\documentstyle[psfig,aps]{revtex}

\def\be{\begin{equation}}
\def\ee{\end{equation}}
\def\bea{\begin{eqnarray}}
\def\eea{\end{eqnarray}}

\def\IR{{\hbox{{\rm I}\kern-.2em\hbox{\rm R}}}}
\def\IB{{\hbox{{\rm I}\kern-.2em\hbox{\rm B}}}}
\def\IN{{\hbox{{\rm I}\kern-.2em\hbox{\rm N}}}}
\def\IC{\,\,{\hbox{{\rm I}\kern-.59em\hbox{\bf C}}}}
\def\IZ{{\hbox{{\rm Z}\kern-.4em\hbox{\rm Z}}}}
\def\IP{{\hbox{{\rm I}\kern-.2em\hbox{\rm P}}}}
\def\IH{{\hbox{{\rm I}\kern-.4em\hbox{\rm H}}}}
\def\ID{{\hbox{{\rm I}\kern-.2em\hbox{\rm D}}}}
\def\II{{\hbox{\rm I}\kern-.2em\hbox{\rm I}}}

\begin{document}
\title{\begin{flushright}
{\small hep-th/9902170}
\end{flushright}
Charged AdS Black Holes and Catastrophic Holography}
\author{Andrew Chamblin$^a$, Roberto Emparan$^b$, Clifford
V. Johnson$^c$ and Robert C. Myers$^d$} \author{} \address{$^a$
D.A.M.T.P., Silver Street, Cambridge, CB3 9EW, UK.}  \address{$^b$
Department of Mathematical Sciences, University of Durham, DH1 3LE,
UK.\\ Departamento de F{\'\i}sica Te\'orica, Universidad del Pa{\'\i}s
Vasco, Apdo. 644, E-48080 Bilbao, Spain.}  \address{$^c$ Department of
Physics and Astronomy, University of Kentucky, Lexington, KY
40506--0055, USA.}  \address{$^d$Physics Department, McGill
University, Montr\'{e}al, PQ, H3A 2T8, Canada.}  \address{\small
$^a$H.A.Chamblin@dampt.cam.ac.uk, $^b$Roberto.emparan@durham.ac.uk,
$^c$cvj@pa.uky.edu, $^d$rcm@hep.physics.mcgill.ca}

\author{}

\maketitle \begin{abstract} We compute the properties of a class of
charged black holes in anti--de Sitter space--time, in diverse
dimensions. These black holes are solutions of consistent
Einstein--Maxwell truncations of gauged supergravities, which are
shown to arise from the inclusion of rotation in the transverse space.
We uncover rich thermodynamic phase structures for these systems,
which display classic critical phenomena, including structures
isomorphic to the van der Waals--Maxwell liquid--gas system. In that
case, the phases are controlled by the universal ``cusp'' and
``swallowtail'' shapes familiar from catastrophe theory. All of the
thermodynamics is consistent with field theory interpretations {\it
via} holography, where the dual field theories can sometimes be found
on the world volumes of coincident rotating branes.
\end{abstract}

\section{Introduction} 

There is evidence that there is a
correspondence\cite{juan,gubkleb,edads} between gravitational physics
in anti--de Sitter space--time and particular types of conformal field
theory in one dimension fewer. This duality is a form of
``holography''~\cite{hologram} and a part of the correspondence
operates by identifying the field theory as living on the boundary of
anti--de Sitter (AdS) space--time.

To be more precise, AdS$_{n+1}$$\times {\cal M}^m$ is the space--time
of interest, and there is some $(n{+}m{+}1)$--dimensional theory of
gravity compactified on it. The manifold ${\cal M}^m$ can be an
$m$--sphere, $S^m$.  The corresponding field theory is an
$n$--dimensional conformal field theory living on a space with the
topology of the boundary of AdS$_{n+1}$. The isometries of the
manifold ${\cal M}^m$ appear as global symmetries of the field theory:
R--symmetries if the theory is supersymmetric.

This particular form of duality between gravity and field theory is
certainly intriguing. The large $N$ limit (where~$N$ is the rank of
the $SU(N)$ gauge group for the four dimensional Yang--Mills field
theory, with appropriate generalizations for other dimensions) of the
field theory ---at strong 't Hooft coupling--- corresponds to
classical supergravity. As pointed out in ref.~\cite{cejm}, following
the observations in ref.~\cite{edads}, the old program of
semi--classical quantum gravity finds a new lease on life in this
setting, as computations such as those performed with gravitational
instantons (at least in AdS) should have natural field theory
interpretations.

In this paper, we study the thermal properties of Einstein--Maxwell
AdS (EMadS) charged black holes, and find behaviour consistent with
field theory interpretations. We do this for arbitrary dimensions
(greater than 3--- see section~\ref{conclude} for comments on $D{=}3$)
and determine the thermal phase structure of the corresponding field
theories. The cases of AdS$_4$, AdS$_5$ and AdS$_7$ are particularly
interesting of course, as they correspond to the theories found on the
world volumes of M2--, D3--, and M5--branes, respectively. The
D3--brane case is $D{=}4$, ${\cal N}{=}4$ supersymmetric $SU(N)$
Yang--Mills theory, while the others are exotic superconformal field
theories~\cite{exotic}. We remark on the field theory interpretation
of our new results in the light of holography.

This paper is also of relevance beyond mere considerations of
holography. Some of the black hole solutions and their properties
(thermodynamic or otherwise) are presented here for the first
time\footnote{The thermodynamics of
Reissner--Nordstrom--anti--deSitter black holes in four dimensions has
been studied, with a slightly different focus, in
ref.\cite{jorma}.}. In particular, the Lagrangian action calculations
and subsequent determination of the phase structure are presented in
their entirety here.

In section~\ref{sec:EMADS}, we present an ansatz for obtaining the
Einstein--Maxwell truncation of gauged AdS supergravity with
appropriate compactifications of $D{=}11$ supergravity on $S^7$, and
$D{=}10$ type IIB supergravity on $S^5$.  In the planar or
infinite--volume limit, the charged black holes in
Einstein--Maxwell--Anti--deSitter correspond to the near horizon
limits of rotating M2-- and D3--branes. In section~\ref{sec:holes}, we
display the solutions and note some of their properties.  The
computation of the action of the solutions using a Euclidean section
is performed in section~\ref{sec:action}, and their thermodynamic
properties are uncovered in section~\ref{sec:thermo}.

As the Einstein--Maxwell--Anti--deSitter truncation is naturally
associated to rotating branes, (at least in the case of EM--AdS$_4$
and EM--AdS$_5$, see section~\ref{sec:EMADS}) it is very natural to
suppose that there is an associated dual field theory, arising on the
world-volume of some branes. These would be the familiar conformal
field theories ---the $D{=}4$, ${\cal N}{=}4$ Yang--Mills theory (for
coincident D3--branes) and the conformal field theory on the
world--volume of coincident M2--branes.  The case of EM--AdS$_7$ ({\it
i.e.,} without additional scalars) is not related to a rotating--brane
truncation of the AdS$_7{\times}S^4$ gauged supergravity (because
$S^4$ is even dimensional) and so we cannot declare that the dual
field theory is the theory on the world--volume of a rotating
M5--brane. However, we regard AdS holography as a phenomenon which
exists independently of string-- and M--theory
contexts\cite{edads,cejm}. Hence, in other dimensions beyond $D{=}4$
and 5, we expect that there is a dual theory. In particular, for
EM--AdS${}_7$ the dual field theory is probably a close cousin of the
M5--brane theory.

The dual field theories have their supersymmetry (if they had any to
start with) broken due to coupling to a global background $U(1)$
current (as well as turning on a non--zero temperature). The CFT is in
a thermal ensemble for which a certain $U(1)$ charge density has also
been ``turned on''. In the ensemble, the expectation value of this
charge breaks the global $SO(m{+}1)$ R--symmetry of the CFT. On the
AdS side, the electromagnetic charge carried by the black holes is in
the same $U(1)$ of corresponding $SO(m{+}1)$ gauge group.

We find very interesting phase structures at intermediate temperatures
(in finite field theory volume), as a result of studying two
complementary thermodynamic ensembles: We study thermodynamic
ensembles with fixed background potential ---in which case the
background is AdS with a constant fixed potential--- and we also study
a fixed localized charge ensemble, for which the background is an
extremal black hole with that charge.  

In all cases, at sufficiently high temperature the physics is
dominated by highly non--extreme black holes, and we therefore recover
the ``unconfined'' behaviour characteristic of the associated field
theories\cite{edads,edadsii}.  The finite horizon size of the black
holes controls the behaviour of the expectation value of spatial
Wilson lines accordingly, yielding the area law behaviour, as follows
from ref.\cite{edadsii}.

At intermediate temperatures, in the fixed charge ensemble, the
presence of charge allows a new branch of black hole solutions to
modify the qualitative phase structure in the low charge regime,
resulting in a very interesting phase structure about which we will
have more to say later in this section.

Intriguingly, as there is an extremal ---but non--supersymmetric---
black hole with non--zero entropy even at zero temperature, we must
conclude something interesting about the field theory in the presence
of the global background $U(1)$ current: There must still be at $T{=}0$ a
large number of states (with the given charge) available to the field
theory in order to generate this entropy.  For the case where we hold
the potential ({\it i.e.,} {\it not} the charge) fixed, we do not
expect that this is the ground state, because the extremal black hole
can decay into Kaluza--Klein particles, leaving AdS. This is
because the extremal black hole is not supersymmetric\footnote{There
do exist supersymmetric solutions here, but they all have naked
singularities\cite{romans,lee}. Furthermore, due to a lack of
horizons, their Euclidean section does not permit a definite
temperature to be defined. These solutions are nevertheless
interesting. The fact that they do not play a role in the phase
structure which we examine here does not mean that they may not have a
role in other AdS physics and thus ultimately be relevant to the dual
field theory.}.

This subtlety does not arise in the standard Gibbons--Hawking calculus
of the thermodynamics of black holes ---which we use here--- because
the calculations are not sensitive to the ability of the black holes
to emit charged particles. 

That the extremal black hole can decay by emitting charged
Kaluza--Klein particles here follows from the fact that the charge
descends from  rotation in higher dimensions. There are well--known
classical processes for reducing the rotation of objects like black
holes by scattering\cite{penrose}, and therefore in the context of
quantum field theory, one has the analogous processes of emission in
superradiant modes\cite{xx}. The same superradiant emission was
considered in the context of charged black holes in
ref.\cite{yy,zz}. Thus one should expect the extremal black hole in
the EMadS truncation to decay {\it via} such supperradiant emission. Of
course, the usual thermal Hawking radiation may also tend to discharge
nonextremal black holes\cite{zz,hawkm,uu}.  In the fixed potential ensemble,
as the charge of the black hole is allowed to fluctuate while it is in
contact with the thermal reservoir, superradiant and Hawking emission
processes can occur to reduce the charge of the black hole, allowing
it to decay back to AdS (plus charge\footnote{Note that the same
thought experiments which do not allow the Penrose process to produce
a naked singularity\cite{naked} will work here also, preventing us
from connecting to the set of solutions representing naked
singularities mentioned above, which do not have the standard thermal
treatment.}). However, in the fixed  charge thermodynamic
ensemble (with varying potential), the extremal black hole is expected
to be the long--lived state at zero temperature.

Translating the formula for the entropy to the field theory we find,
for example, that the four dimensional Yang--Mills theory (in the
presence of the global background $U(1)$ current) has a
zero--temperature entropy which goes like $S{\sim}{\bar Q}$ for large
black holes, where $\bar Q$ measures the total charge in units of the
minimal charge of Kaluza--Klein excitations ({\it i.e.,} $1/l$), and
is proportional to the volume, $V_3$, of the field theory.  Notice that
the result for the four dimensional field theory is consistent with
confinement at $T{=}0$, as the result is independent of $N$.
Confinement also follows from the fact that at $T{=}0$, the Euclidean
section of the solution has no bolt, and therefore temporal Wilson
lines will always be homotopic to zero, and therefore have zero
expectation value. Meanwhile, spatial Wilson lines cannot interact
with the horizon to produce an area law dependence, because at
extremality the horizon recedes infinitely far away down a
Bertotti--Robinson throat.

\begin{figure}[ht]
\hskip0.5cm
\psfig{figure=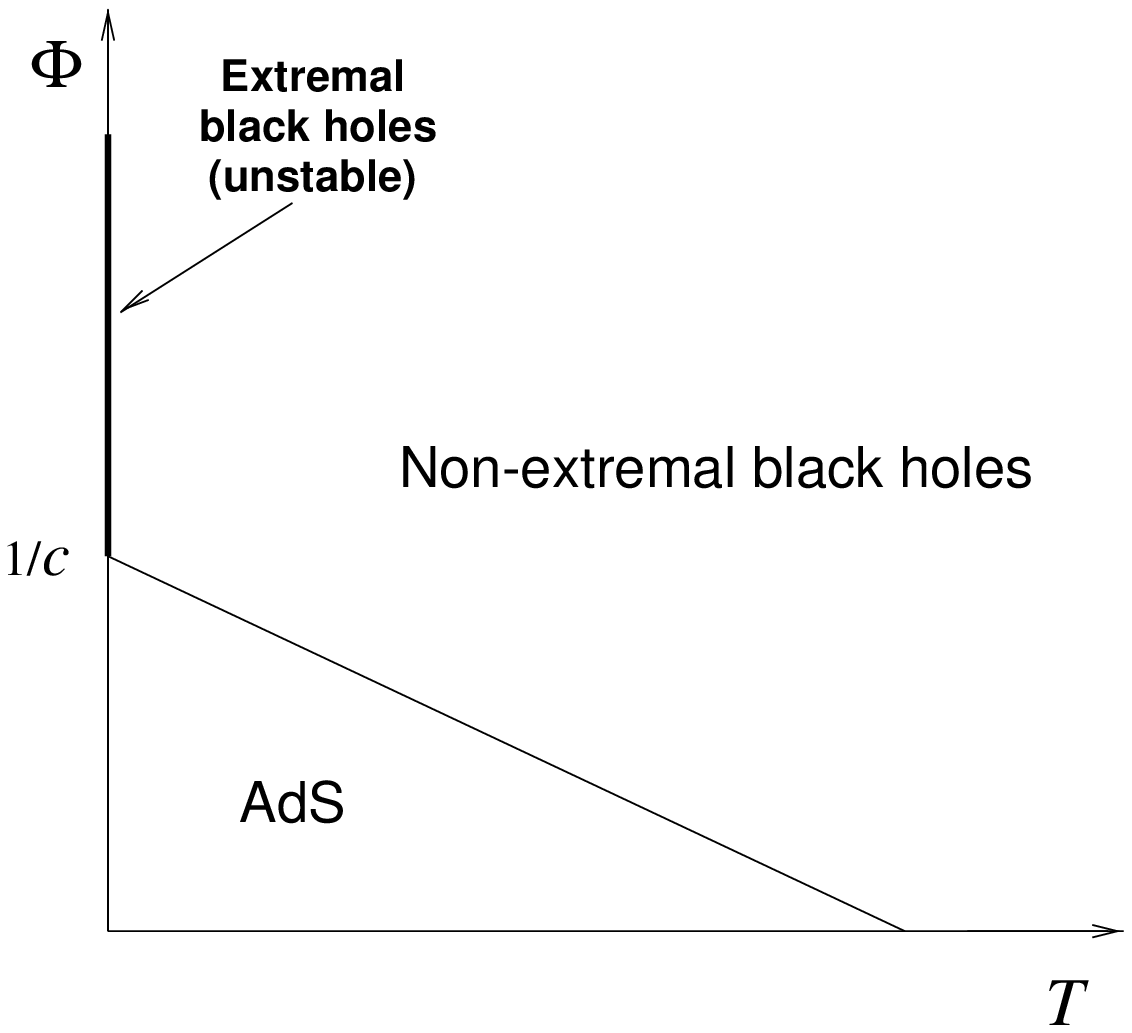,height=2.5in}
\hskip2.0cm
\psfig{figure=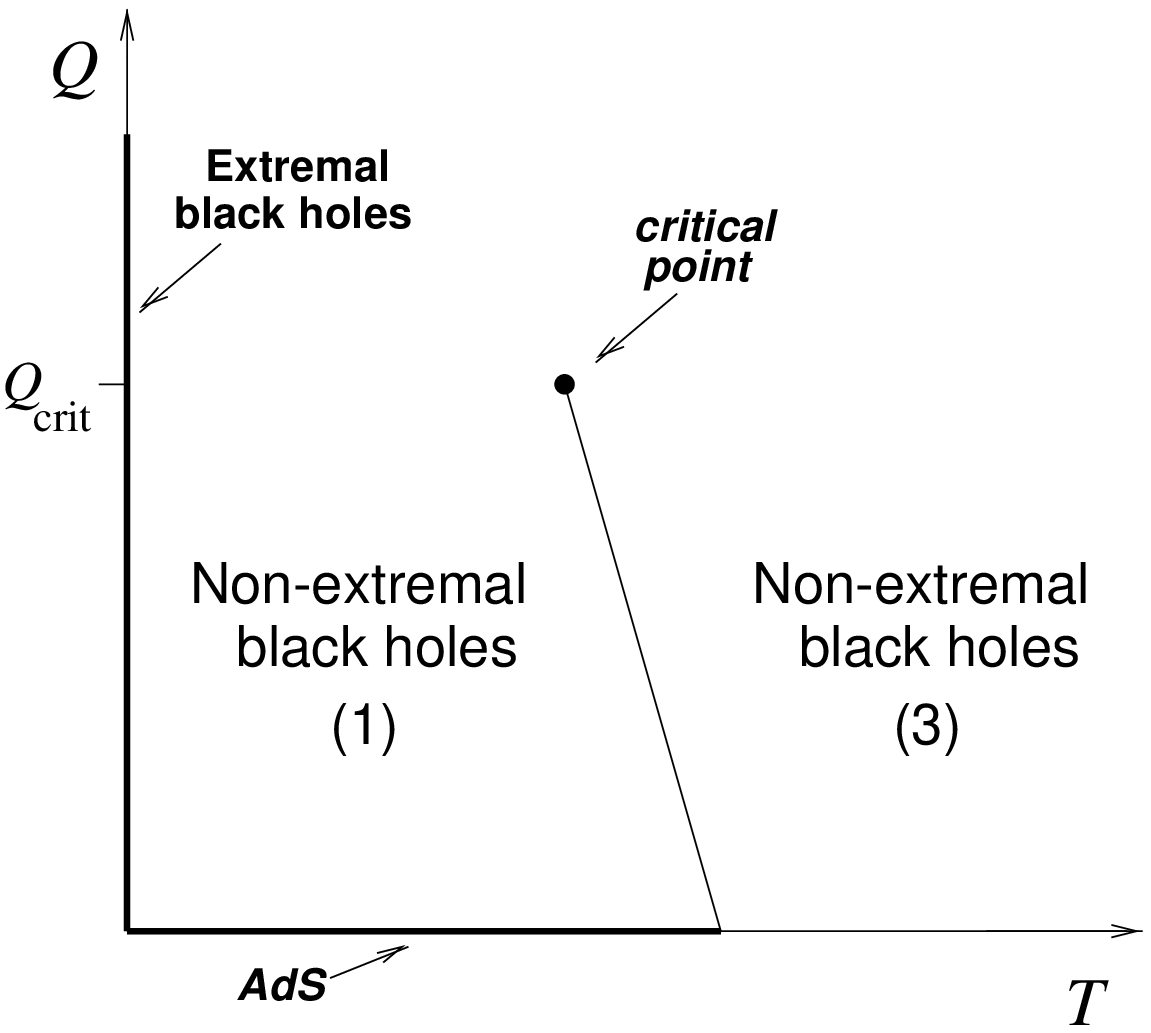,height=2.5in}
\caption{A summary of the phase structure of the fixed potential
(left) and fixed charge (right) thermodynamic ensembles. The $T{=}0$
line gives extremal black holes, although only in the fixed charge
case do they not decay into AdS. The $Q{=}0$ line is the Hawking--Page
system of uncharged black holes. (Other labeling is explained in
sections~\ref{sec:thermo} and~\ref{cats}.)}
\label{fig:phases}
\end{figure}

The phase structure which we obtain in each thermodynamic ensemble is
summarized in figure~\ref{fig:phases}.

The astute reader will recognize the figure on the right as the
classic phase diagram of the liquid--gas system! To translate, our $Q$
is like the temperature $T$ of the fluid while $1/T$ is like the
pressure $P$. The non--extreme black holes of type (1) (``small'') and
(3) (``large'') (see section~\ref{sec:action} and \ref{sec:thermo} for
explanation) are like the liquid phase and the gaseous phase,
respectively. The critical line (``vapour pressure curve'') represents
the place at which a first order phase transition between the liquid
and gas occurs. As is well known, there is a critical temperature at
which the vapour pressure curve terminates, representing the fact that
above a critical temperature, one can convert a liquid to a gas
continuously. This translates here into a critical charge above which
the two types of black hole can be continuously converted into one
another with no discontinuity in their size.

That this system (first modeled by van der Waals\cite{van}, with a
crucial modification by Maxwell) appears in this AdS black hole
thermodynamics is fascinating, and would not have been possible (at
least in this way) without the presence of the extra branches of
solutions which appear when there is negative cosmological
constant. We discuss this further in sections~\ref{sec:thermo}
and~\ref{sec:action}. Further fascination may be found in the fact
that the explicit shape of the free energy surface (as a function of
$Q$ and $T$) is that of the classic ``swallowtail'' catastrophe,
familiar from the study of bifurcations\cite{catastrophe}. The control
surface of the ``cusp'' catastrophe also appears, which (of course)
follows from the well known fact that it is the shape of the van der
Waals equation of state, viewed as a surface in $P,V,T$ space.

That these shapes appear in this context suggests that there is some
exciting universality to be explored here: Catastrophe theory is
largely a classification of the possible distinct types of bifurcation
shapes that can occur in a wide variety of complex systems. This
classification (which, for the common ``elementary'' cases is of
A--D--E type) is equivalent to the (perhaps more familiar)
classfication of singularities\cite{arnold}. It is of considerable
interest to discover just what circumstances might give rise to the
other members of the classification. Recalling that this all
translates {\it via} holography into properties of a dual field
theory, we would learn a great deal about universal phase structures
which can occur there also.

\section{Einstein--Maxwell--AdS from spinning branes}
\label{sec:EMADS}

Physics near the horizon of supergravity branes can be described in
terms of spontaneous compactification of supergravity.  In the case of
non--dilatonic branes ---which will be the focus of the paper--- when
the compactification takes place on a {\it round} $m$--sphere the low
energy degrees of freedom are described by an effective theory of
Einstein gravity with a negative cosmological constant coupled to
$SO(m{+}1)$ gauge fields. The Schwarzschild--Anti--deSitter black hole
solutions of this theory have been used in the context of the AdS/CFT
correspondence to infer thermal properties of the dual field
theories~\cite{edads,edadsii}.

A natural extension of this program is to study AdS black holes which
are charged under a subgroup of the gauge symmetry of the gauged
supergravity. Solutions of Einstein--Maxwell--anti--deSitter in some
dimensions are known, but in the context of string/M--theory, it is
also interesting to determine how to make a truncation of the type~IIB
supergravity, or of 11 dimensional supergravity, which gives the EMadS
effective action. In other words, we must make certain
higher--dimensional choices which will result in the removal of the
generic coupling of the $F^2$ term to scalars resulting from the
Kaluza--Klein reduction.

Amusingly, one simple way to introduce (gauge) charge on the black
holes is by simply spinning ---or twisting--- the transverse (angular)
sphere that becomes the compact space.  Decoupling of the scalars is
accomplished by choosing the spins in a maximally symmetric way. To be
concrete, take ten dimensional IIB supergravity, with the metric
ansatz
\begin{equation}
ds^2_{10}=g^5_{\mu\nu}dx^\mu dx^\nu +l^2\sum_{i=1}^3 [d\mu_i^2+\mu_i^2
(d\varphi_i+\frac{2}{\sqrt{3}}A_\mu dx^\mu)^2],
\label{ansatz}
\end{equation}
where $g^5_{\mu\nu}$ is a five-dimensional metric, $\mu,\nu=0\dots 4$,
the variables $\mu_i$ are direction cosines on $S^5$ (and therefore
are not independent, $\sum_{i=1}^3 \mu_i^2=1$ ---we follow the
notation of\cite{nn}), and the $\varphi_i$ are rotation angles on
$S^5$. 
The ansatz for the RR 5-form field strength has 
``electric'' components 
\begin{equation}
F^{(5)}_{e} = -{4\over l}\varepsilon^{(5)} + {l^3\over\sqrt{3}}
\sum_{i=1}^3d\mu_i^2 d\phi_i\,*^5\!dA,
\label{forms}
\end{equation}
while the dual ``magnetic'' components are given by
$F^{(5)}_m{=}*\! F^{(5)}_e$.  In eqn.~(\ref{forms}),
$\varepsilon^{(5)}$ is the volume form on the reduced five-dimensional
space, and $*^5$ denotes Hodge duality on this space.

The parameter $l$ measures the size of the $S^5$ and is given by the
flux of the 5--form field across the $S^5$. Notice that a component
$A_t$ in the time direction is interpreted as rotation of the $S^5$ in
its three independent rotation planes, in equal amounts. Components in
the spatial direction would instead be `twists.' For the sake of
brevity, and since in this paper we will be mainly considering $A_t$
components\footnote{In any event for $d\ge5$, one cannot define
magnetic (vector) charges on the black holes.}, we will refer
collectively to them as ``rotations''.

With this ansatz~(\ref{ansatz}), the effective action in the five
non--compact dimensions becomes
\begin{equation}
I=-{1\over 16 \pi G_5}\int d^5x\sqrt{-g^5}\left[ R+{12\over l^2} 
-{l^2}F^2-{l^3\over 6\sqrt{3}}\epsilon^{\mu\alpha\beta\gamma\delta}
A_\mu F_{\alpha\beta}F_{\gamma\delta} \right].
\end{equation}
This is precisely the Einstein--Maxwell--Anti--deSitter (EMadS)
effective action we seek, with a Chern--Simons term. The latter is
indeed required by supersymmetry in ${\cal N}{=}2$ five dimensional
gauged supergravity~\cite{romans}, whose bosonic sector is precisely
described by the action above. Note that the gauge coupling is
proportional to $\sqrt{G_5}/l$.

The AdS$_5{\times}S^5$ gauged supergravity theory in five dimensions
has an $SO(6)$ gauge symmetry, associated with the group of isometries
of $S^5$. This is the R--symmetry group of the dual four dimensional
${\cal N}{=}4$ superconformal Yang--Mills field theory living on the
D3--branes from which this near--horizon geometry arose.  The above
spinning compactification corresponds to introducing rotation in the
diagonal $U(1)$ of the maximal Abelian subgroup $U(1)^3$.
Correspondingly, there must be a dual field theory to the EMadS
truncation, which is simply the field theory on the world--volume of
the rotating brane. From the field theory point of view, the rotation
corresponds to considering states or ensembles in which the dual
global $U(1)$ current (a subgroup of the $SO(6)$ R--symmetry group)
has a nonvanishing expectation value. Studying EMadS gravity and its
solutions will therefore be equivalent to studying properties of the
conformal field theory in the presence of this background
current\footnote{A more general action can be constructed, that
contains three $U(1)$ vector fields, each associated with the three
different independent rotations of $S^5$, and two scalars that,
roughly, measure the relative sizes of the distortions of the $S^5$
caused by rotation. For simplicity, we will restrict ourselves to the
case where all three rotations have the same magnitude, since it is
only in this case that the scalars decouple and we find EMadS
gravity. This framework provides the cleanest interpretation in terms
of the dual CFT, since the number of spin parameters or charges
precisely matches the number of field theory operators which are
``excited''. See refs.~\cite{14.1,14.2,14.3} for discussion of more
general actions and solutions related to this.}.

A similar construction can be obtained by starting from eleven
dimensional supergravity.  The compactification in this case is
equivalent to focusing on the near horizon region of M2--branes. In
this case, take
\begin{equation}
ds^2_{11}=g^4_{\mu\nu}dx^\mu dx^\nu + 4l^2\sum_{i=1}^4 [d\mu_i^2+\mu_i^2
(d\varphi_i+A_\mu dx^\mu)^2],
\end{equation}
leading to the AdS$_4$ theory with a Maxwell term
\begin{equation}
I=-{1\over 16 \pi G_4}\int d^4x\sqrt{-g^4}\left[ R+{6\over l^2} 
-4l^2F^2\right].
\end{equation}
The reduction ansatz for the 4-form field strength is
\begin{equation}
F^{(4)}= {3\over l} \varepsilon^{(4)} +4l^2 \sum_{i=1}^4 d\mu_i^2
d\phi_i\,*^4\!dA,
\end{equation}
where $\varepsilon^{(4)}$ is the volume form on the reduced
four--dimensional space, and $*^4$ denotes Hodge duality on this
space.

Chern--Simons terms are absent in four dimensions. Appropriate
inclusion of fermions leads to four dimensional ${\cal N}{=}2$ gauged
supergravity. The more general $U(1)^4$ theory with four independent
gauge fields ({\it i.e.}, four different rotation parameters) 3
scalars and ${\cal N}{=}8$ supersymmetry, as well as its black hole
solutions, have been recently studied in ref.~\cite{duffliu}.

We note here that there is no analogous construction for the
AdS$_7{\times}S^4$ gauged supergravity theory. This is because~$S^4$
is even dimensional and therefore we cannot have a symmetric split
between $U(1)$ rotations, as $SO(5)$ does not have an even torus for
its Cartan subalgebra. This means that we cannot relate the physics of
the black hole solutions (which we write later) of the EM--AdS$_7$
system to the physics of rotating M5--branes of eleven dimensional
supergravity. Nevertheless, as AdS holography is a phenomenon which is
expected to exist independently of string or M--theory realizations,
we expect that the physics does have a holographic interpretation in
terms of a field theory closely related to that which resides on
M5--brane world--volumes.

\section{Charged black holes in Anti--de Sitter space--time}
\label{sec:holes}

The black hole solutions of the above supergravity theories in
$D{=}4,5$ were originally studied in the past in refs.~\cite{romans}
and~\cite{lee}---more recent investigations appear in
refs.~\cite{14.1,duffliu}. As we have seen in the previous section, such
theories can be regarded as compactifications of the type IIB and
$D{=}11$ supergravities, where the gauge symmetry groups of the gauged
supergravities are broken by a specific choice of rotation planes in
the transverse compact spheres.  Given these considerations, it is
natural to study the Reissner--Nordstrom--anti--deSitter (RNadS) black
holes within the context of the AdS/CFT correspondence.

Even if the bosonic Einstein--Maxwell--Anti--deSitter (EMadS) theories
admit supersymmetric extensions only in certain dimensions, it is easy
and convenient to perform the analysis of their black hole solutions
for arbitrary dimension. For space--time dimension $n{+}1$, the action
can be written as\footnote{We rescale the gauge field $A_\mu$ so as to
absorb the prefactors in the action.}
\begin{equation}
I = -\frac{1}{16{\pi}G} \int_{M} d^{n+1}x \sqrt{-g} \left[R - F^2 +
\frac{n(n-1)}{l^2}\right],
\label{actionjackson}
\end{equation}
with ${\Lambda}{=}{-}\frac{n(n-1)}{2l^2}$  the cosmological constant
associated with the characteristic length scale $l$.  
Then the metric on RNadS may be written in 
static coordinates as
\begin{equation}
ds^2 = -V(r)dt^2 + \frac{dr^2}{V(r)} +r^{2}d{\Omega}^2_{n-1},
\end{equation}

\noindent where $d{\Omega}^2_{n-1}$ is the metric on the round
unit $(n{-}1)$--sphere, and the function $V(r)$ takes the form
\begin{equation}
V(r) = 1 - \frac{m}{r^{n-2}} + \frac{q^2}{r^{2n-4}} + \frac{r^2}{l^2}.
\end{equation}

\noindent Here, $m$ is related to the ADM mass of the hole, $M$
(appropriately generalized to geometries asymptotic to
AdS\cite{adhh}), as
\begin{equation}
M={(n-1)\omega_{n-1}\over 16\pi G}m,
\end{equation}
 where $\omega_{n-1}$ is the volume of the unit $(n{-}1)$--sphere.
The parameter
$q$ yields the charge
\begin{equation}
Q=\sqrt{2(n-1)(n-2)}\left({\omega_{n-1}\over 8\pi G}\right)q,
\label{thecharge}
\end{equation}
of the 
(pure electric) gauge potential, which is
\begin{equation}
A=\left(-{1\over c}{q\over r^{n-2}}+\Phi\right)dt,
\label{pure}
\end{equation}
where
\begin{equation}
c=\sqrt{2(n-2)\over n-1},
\end{equation}
and $\Phi$ is a constant (to be fixed below). If $r_+$ is the largest
real  positive root of $V(r)$, then in order for this RNadS metric to
describe a charged black hole with a non--singular horizon at $r{=}r_+$, the
latter must satisfy 
\begin{equation}\label{extbound}
\left({n\over n-2}\right) r_+^{2n-2}+l^2 r_+^{2n-4} \geq q^2 l^2.
\end{equation}
Finally, we choose
\begin{equation}
\Phi={1\over c}{q\over r_+^{n-2}},
\end{equation}
which then fixes $A_t(r_+){=}0$. The physical significance of the
quantity $\Phi$, which plays an important role later, is that
it is the electrostatic potential difference between the
horizon and infinity.

If the inequality in eqn.~(\ref{extbound}) is saturated the horizon is
degenerate and we get an extremal black hole. This inequality imposes
a bound on the black hole mass parameter of the form $m{\geq}m_e(q,l)$.
In the cases where the theory admits a supersymmetric embedding one
could naively expect to approach a supersymmetric state as we saturate
this mass bound.  However, the bound that results from the
supersymmetry algebra is instead~\cite{romans,lee}: $m{\geq}2q$,
with the $m{=}2q$ solution being a BPS state\footnote{In $D{=}4$, where the
black hole can have magnetic charge $q_m$, there is a magnetic (or
dyonic) BPS solution as well \cite{romans} with $m{=}0$,
$q_m{=}\pm l/2$.}.  Now, it is easy to see that the mass of the
extremal black hole, $m_e$ is, for finite $l$, always strictly larger
than $2q$ and therefore the extremal solution is non--supersymmetric.
On the other hand, for the supersymmetric solution one has
\begin{equation}
V(r)=\left(1-{q\over r^{n-2}}\right)^2 +{r^2\over l^2},
\end{equation}
which is strictly positive everywhere and therefore one finds a naked
curvature singularity at $r{=}0$. In fact, all the solutions violating
the bound~(\ref{extbound}) are nakedly singular.

In the context of the AdS/CFT correspondence it is interesting to
consider the limit where the boundary of AdS$_{n+1}$ is $\IR^n$
instead of $\IR{\times}S^{n-1}$ as was the case above. This can be
regarded as an ``infinite volume limit'', with particular relevance to
the discussion of the dual field theory. It should be noted that the
existence of black hole solutions in this limit is possible only due
to the presence of a negative cosmological constant. In fact, black
holes (and other bolts) in AdS spaces with varied topologies (even
other than spherical and toroidal) have been extensively studied in
recent years~\cite{otherholes}, including in the
M--theory~\cite{roberto}. Here we will only focus on the planar
(toroidal) solutions, which we will obtain by scaling the ``finite
volume'' solutions above, as done in \cite{edadsii}. To this effect,
introduce a dimensionless parameter $\lambda$ (which we will shortly
take to infinity) and set
\begin{eqnarray}
r&&\rightarrow \lambda^{1/n} r,\qquad t\rightarrow \lambda^{-1/n} t,
\nonumber\\ m&&\rightarrow \lambda m, \qquad q\rightarrow
\lambda^{(n-1)/n} q, \label{scaled}
\end{eqnarray}
while at the same time blowing up the $S^{n-1}$ as $l^2
d\Omega_{n-1}^2\rightarrow \lambda^{-2/n}\sum_{i=1}^{n-1}dx_i^2$.  One
finds, after taking $\lambda{\to}\infty$,
\begin{equation}
ds^2=-U(r)dt^2+{dr^2\over U(r)}+{r^2\over l^2}\sum_{i=1}^{n-1}dx_i^2,
\end{equation}
with
\begin{equation}
\label{scaledU}
U(r)={r^2\over l^2}-{m\over r^{n-2}}+{q^2\over r^{2n-4}}.
\end{equation}
For the supersymmetric solution, the scaling is as above except for
the scaling of $m$. To preserve supersymmetry, one must fix $m{=}2q$
and so $m{\rightarrow}m\lambda^{(n-1)/n}$ yielding
\begin{equation}
U(r) = {r^2\over l^2}+{q^2\over r^{2n-4}} .
\end{equation}
Notice that, compared with eqn.~(\ref{scaledU}), the parameter $m$ is
zero in this limit.

The resulting solution can be seen to be supersymmetric as well ({\it
i.e.}, the Killing spinors remain finite in the limit
$\lambda{\rightarrow}\infty$, after appropriate rescaling), and nakedly
singular. In this ``infinite volume'' limit, the solutions asymptote to
AdS space with the horospheric slicing.

These planar solutions can be constructed with the appropriate
decoupling limit\cite{juan} of spinning D3-- or M2--branes, as mentioned
previously. We refer the reader to ref.\cite{14.2} for the details.

\section{Action Calculation}
\label{sec:action}

The study of the Euclidean section ($t{\to}i\tau$) of the solution,
identifying the period, $\beta$, of the imaginary time with inverse
temperature, will define for us the grand canonical thermodynamic
ensemble (for fixed electric potential) or the canonical ensemble (for
fixed electric charge).  We interpret this in terms of immersing the
system into a thermal bath of quanta at temperature $T{=}1/\beta$.
For pure AdS, the background consists of both charged and uncharged
quanta free to fluctuate in the presence of fixed potential
$\Phi$. Later, we consider the fixed $Q$ ensemble. In that case we
localize all of the charge at a specific region and keep it fixed. For
such a background, as AdS with a localized charge is {\em not} a
solution of the EMadS equations, we use the extremal black hole
solution as the background, and retain only neutral quanta in the
thermal reservoir, in order to keep the charged fixed. This makes
sense, even though the extreme limit has zero temperature, since the
Euclidean section has no bolt and so can be assigned an arbitrary
periodicity\cite{gg}. Hence, the metrics and gauge fields can be
matched in the asymptotic region.

With all of this in mind we now turn to the action calculations.

\subsection{Fixed Potential}

With our conventions the full Euclidean action is given by
analytically continuing eqn.~(\ref{actionjackson}),
where, as usual when the space is asymptotically AdS the
Gibbons--Hawking boundary term gives a vanishing contribution. The
boundary terms from the gauge field will vanish if we keep the
potential $A_t$ fixed at infinity. Any possible Chern--Simons term
will not contribute when we restrict ourselves to purely electric
solutions.  Imposing the equations of motion we can eliminate the
factors of $R$ in order to obtain the on--shell action
\begin{equation}
\label{master}
I = \frac{1}{16{\pi}G} \int_{M} d^{n+1}x \sqrt{g} \left[\frac{2F^2}{n-1}
+ \frac{2n}{l^2}\right].
\end{equation}

We obtain for the action (subtracting the AdS background while
remembering to match the geometries of the background and black hole
in the asymptotic region):
\begin{eqnarray}
I=\frac{\omega_{n-1}}{16\pi G l^2}\beta\left(l^2
r_+^{n-2}-r_+^n-\frac{q^2l^2}{r_+^{n-2}}\right)
=\frac{\omega_{n-1}}{16\pi G l^2}\beta\left(l^2
r_+^{n-2}(1-c^2\Phi^2)-r_+^n\right).
\label{actionone}
\end{eqnarray}
Here, $\beta$ denotes the period of the
Euclidean section of the black hole space--time.  Using the usual
formula for the period, $\beta{=}{4\pi}/{V^{\prime}(r_{+})}$, a little
algebra yields the explicit form
\begin{equation}
\label{betaform}
\beta = \frac{4{\pi}l^{2}r_{+}^{2n-3}}{nr_{+}^{2n-2} +
(n-2)l^{2}r_{+}^{2n-4} - (n-2)q^{2}l^{2}}.
\end{equation}

\noindent This may be rewritten in terms of the potential as:
\begin{equation}
\beta = \frac{4{\pi}l^{2}r_+}{(n-2)l^{2}(1-c^2\Phi^2)+nr_+^2}.
\label{betaformtwo}
\end{equation}

\begin{figure}[hb]
\hskip-0.5cm
\psfig{figure=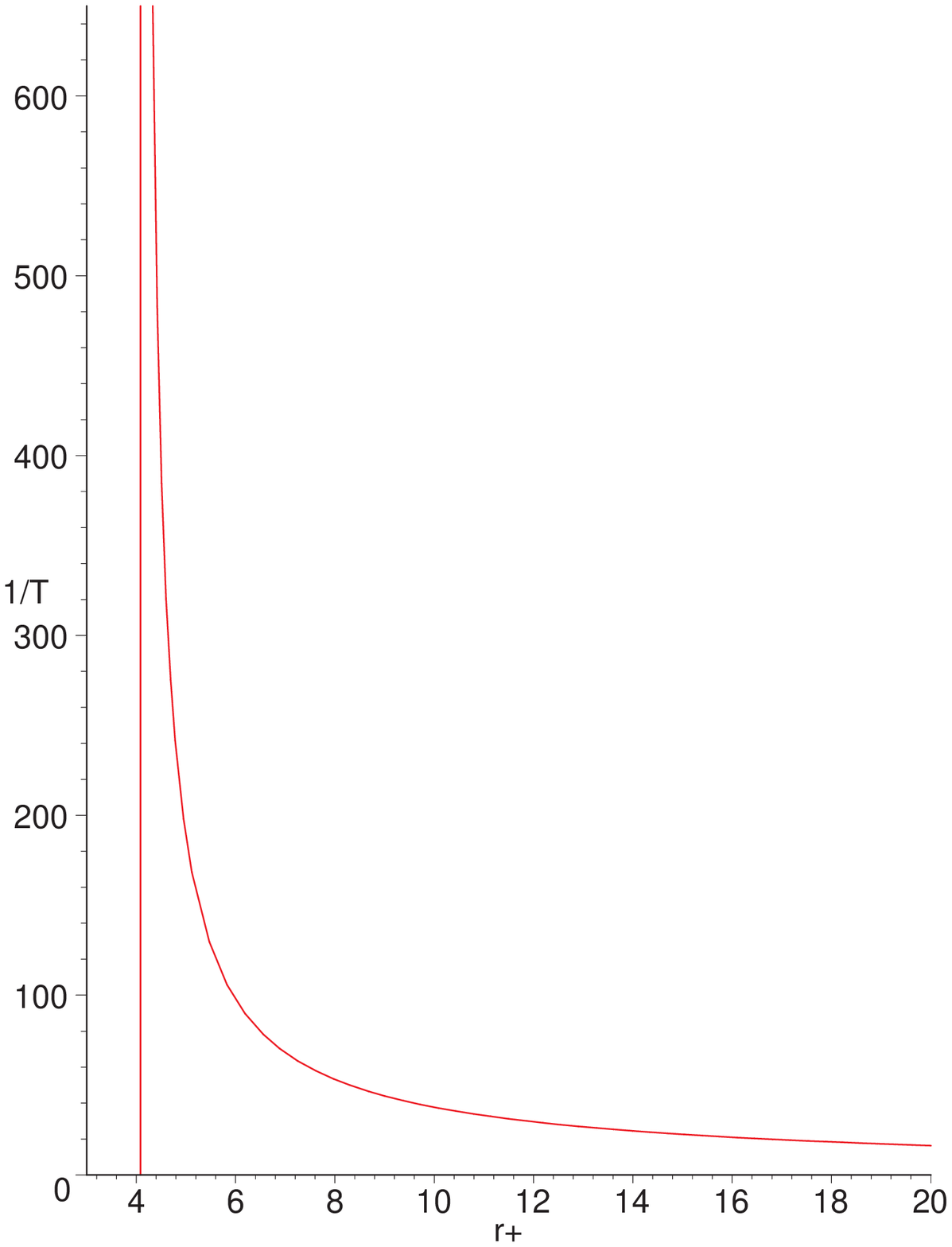,height=3in}
\psfig{figure=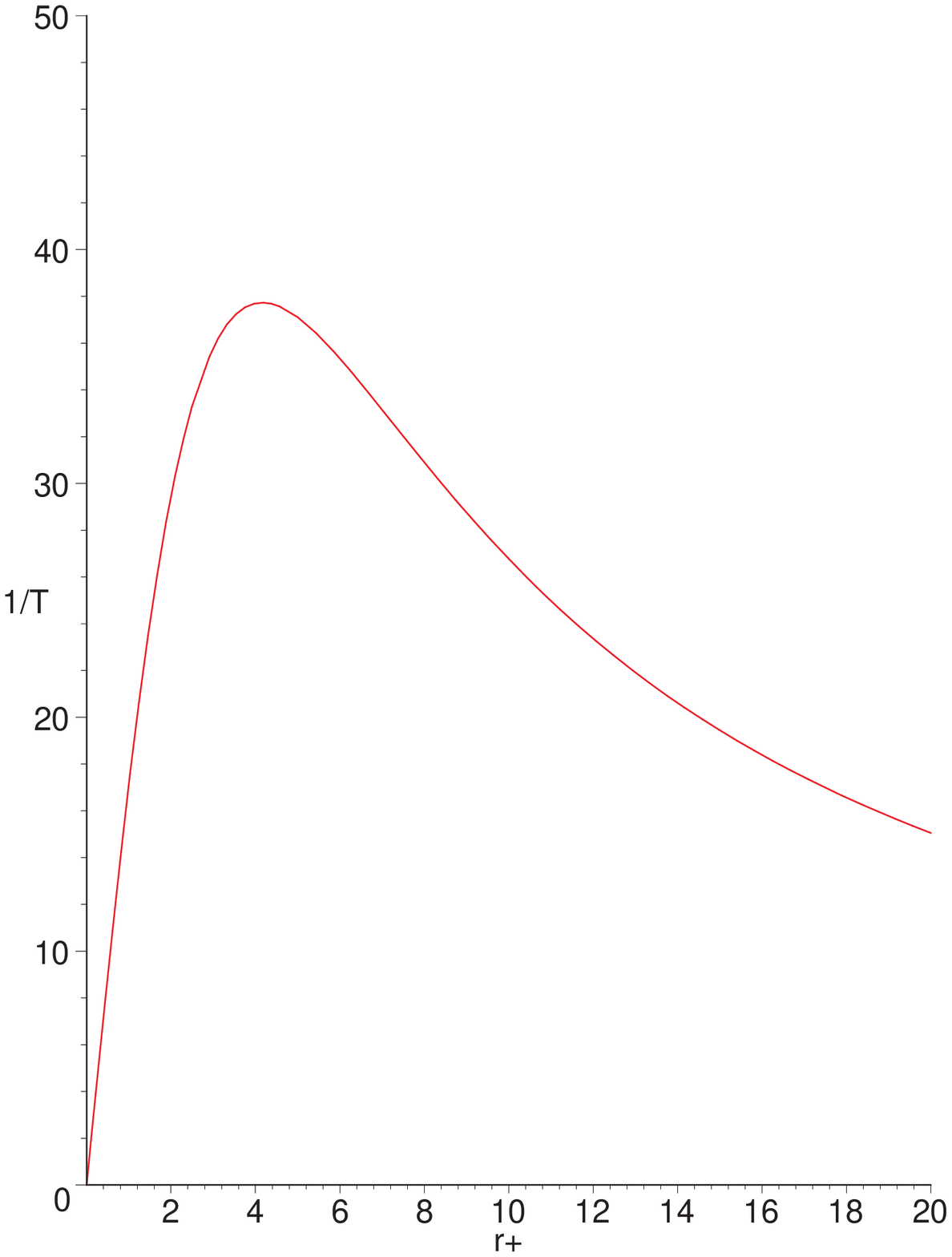,height=3in}
\psfig{figure=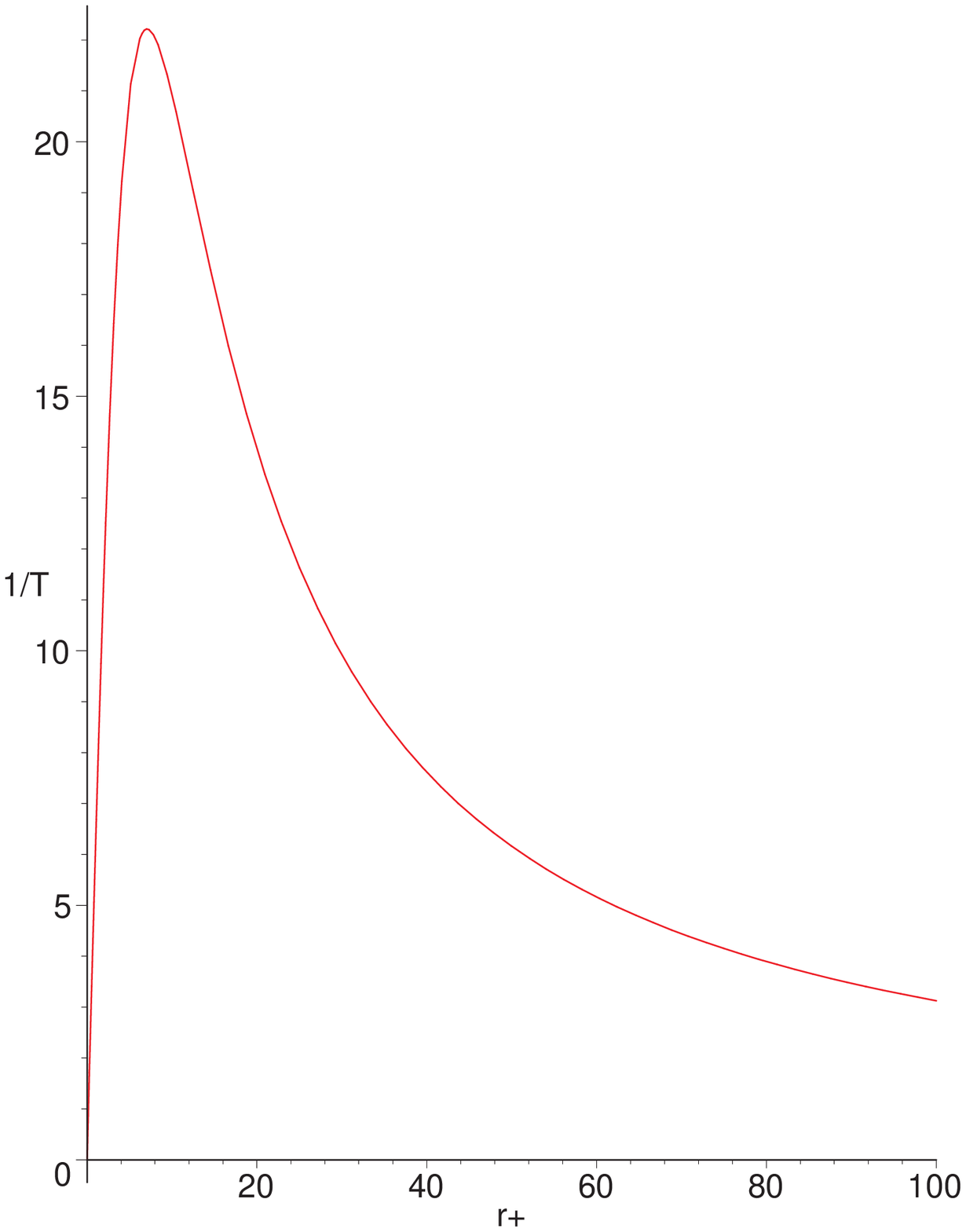,height=3in}
\caption{The inverse temperature {\it vs.} horizon radii, $r_+$, at
fixed potential for $\Phi{\ge}1/c$, $\Phi{<}1/c$, and $\Phi{=}0$
respectively. (The values $n{=}4$, $G{=}1$, $l{=}10$ and
$\Phi=1,0.7,0$ have been used here.)  The divergence in the first
graph (here, shown with a vertical line at $r_e{=}4.08$) is at zero
temperature, where the black hole is extremal. This divergence goes
away for $\Phi{<}1/c$, in general, and the curve is similar to that of
the uncharged situation with zero potential, shown last.}
\label{betaplot3}
\end{figure}

Note that the temperature is zero when the black hole is
extremal. This is because the horizon is degenerate there, and $\beta$
diverges, together with the fact that one can smoothly approach the
extremal limit from non--zero temperature.  From the form of the
equation for $\beta$, it is apparent that there are qualitatively two
distinct types of behaviour, determined by whether $\Phi$ is less than
or greater than the critical value $1/c$. In particular, for
$\Phi{\ge}1/c$, $\beta$ diverges ($T$ vanishes) at
$r_+^2{=}l^2(n-2)(c^2\Phi^2{-}1)/n$, while for $\Phi{<}1/c$, $\beta$ goes
smoothly towards zero as $r_+{\rightarrow}0$. It is instructive to plot
the temperature as a function of horizon radius (size of black hole)
for these two regimes (see fig.~\ref{betaplot3}).

As can be seen from the figure, the regime of large potential ({\it
i.e.,} $\Phi\ge1/c$) has a unique black hole radius associated with
each temperature. We will see later that this branch dominates the
thermodynamics for all temperatures. Meanwhile the small potential
regime has two branches of allowed black hole solutions, a branch with
larger radii and one with smaller. This is qualitatively similar to
the familiar case of the uncharged Schwarzschild black holes analyzed
in ref.~\cite{hawkpage}, (or the structure of the Taub--bolts
discovered in the thermodynamic studies of ref.~\cite{cejm,hawkhunt}),
which is the $\Phi{=}0$ limit of the situation here. Correspondingly, the
smaller branch of holes is unstable, having negative specific
heat. They do not play any role in the physics\footnote{This may be
contrasted with the situation in ref.\cite{14.3}}. (Generally, the sign
of the specific heat for a black hole of radius $r_+$ can be inferred
from the local slope of the $\beta(r_+)$ curve. See also the
discussion in section~\ref{cats}.)

\subsection{Fixed Charge}

If we wish to consider a situation where instead of the potential at
infinity, we fix the charge of the black hole, then the action
(\ref{actionone}) is not appropriate. Upon variation of the gauge
field in the latter action, a boundary term results that vanishes only
if we keep $A_t(\infty){=}\Phi$ fixed. That is, the on--shell action of
the previous subsection is $I[\beta,\Phi]$. If, instead, we want to
keep the charge fixed, then we must add a boundary term to $I$
\cite{haro},
\begin{equation}
\tilde I= I-{1\over 4\pi G}\int d^{n}x \sqrt{h} F^{\mu\nu}n_\mu A_\nu,
\end{equation}
where $n_\mu$ is a radial unit vector pointing outwards
(notice that this boundary term is determined by the terms coming from
the variation of the off--shell action (\ref{actionjackson}), and not
(\ref{master}), which is on--gravity--shell. This distinction is only
relevant for $n{>}3$). Then we get a thermodynamic function $\tilde
I[\beta,Q]$, in terms of the variables we wish to control.

To compute the action for the fixed charge ensemble, using as
background the extremal black hole, we evaluate eqn.~(\ref{master}) for a
black hole of mass $m{>}m_e$ (and radius $r_{+}$), and then subtract
the contribution from the extremal background.  Remembering to match
the geometries of the background and black hole in the asymptotic
region, a straightforward calculation yields the final result
%

\begin{equation}
\tilde I={\omega_{n-1}\beta\over 16\pi G l^2}\left[ l^2 r_+^{n-2}-r_+^n
+{(2n-3)q^2 l^2\over r_+^{n-2}} - {2(n-1)\over n}l^2 r_e^{n-2}
-{2(n-1)^2\over n}{q^2 l^2\over r_e^{n-2}}\right].
\label{actiontwo}
\end{equation}

\begin{figure}[hb]
\hskip-0.5cm
\psfig{figure=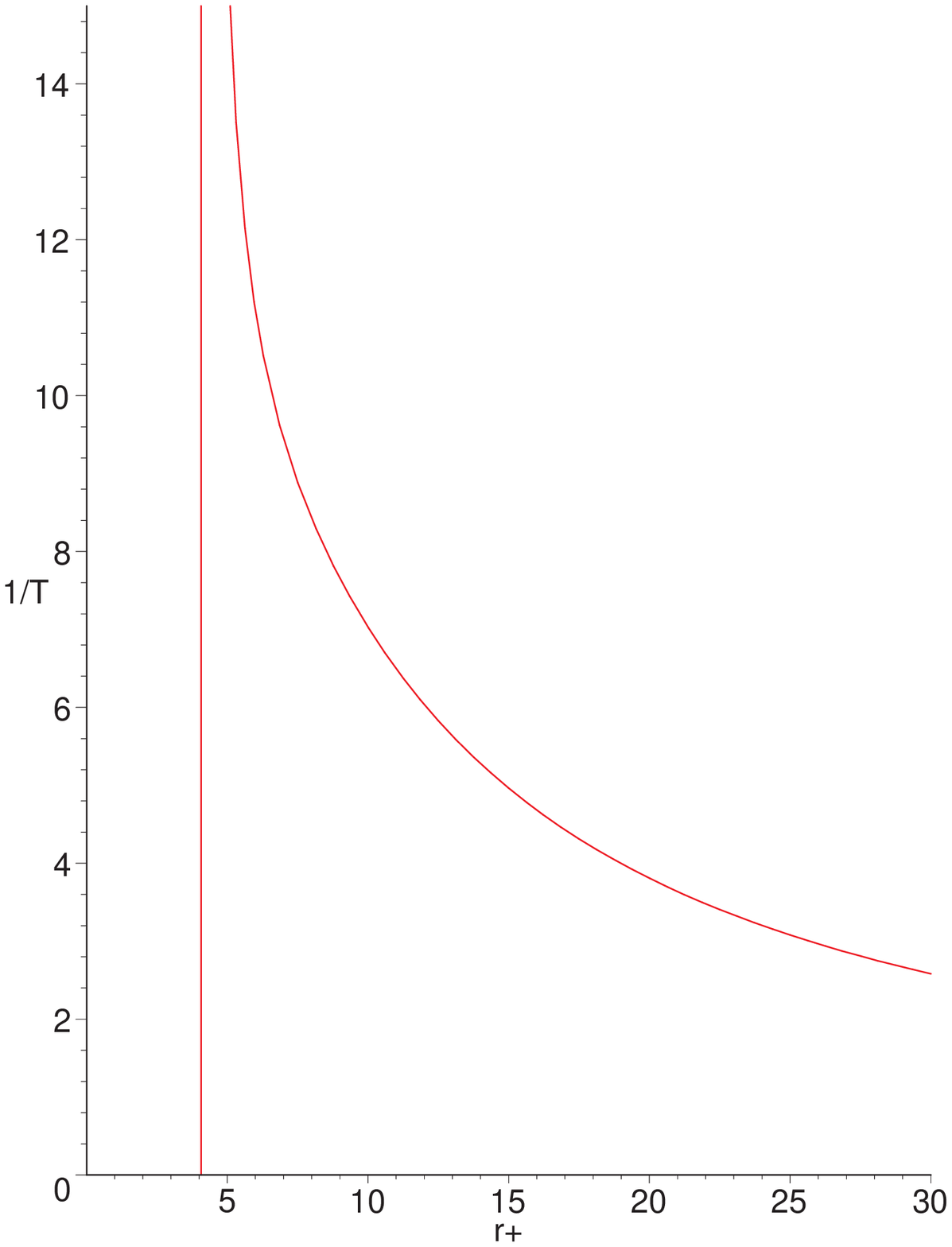,height=3.0in}
\psfig{figure=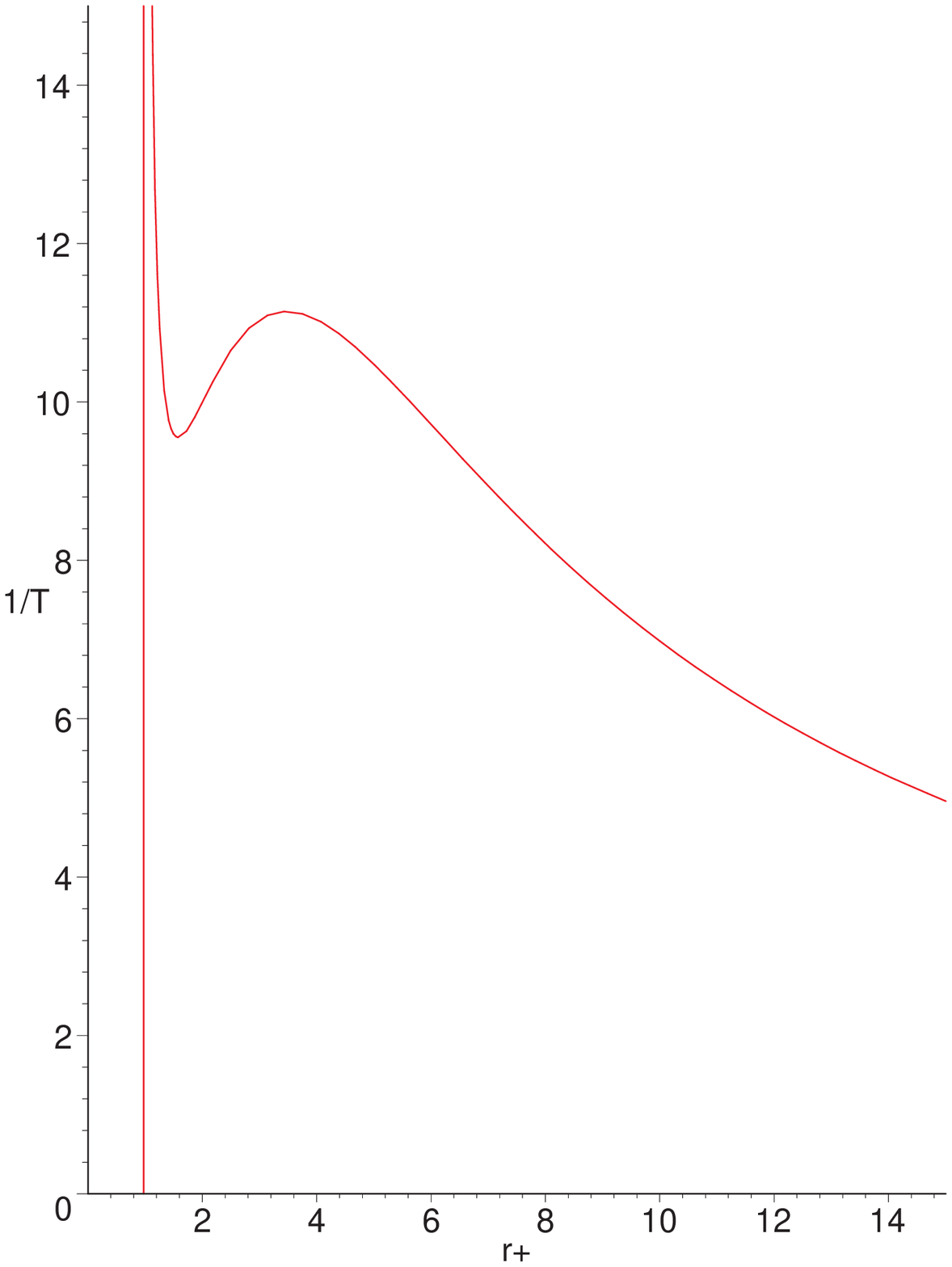,height=3.0in}
\psfig{figure=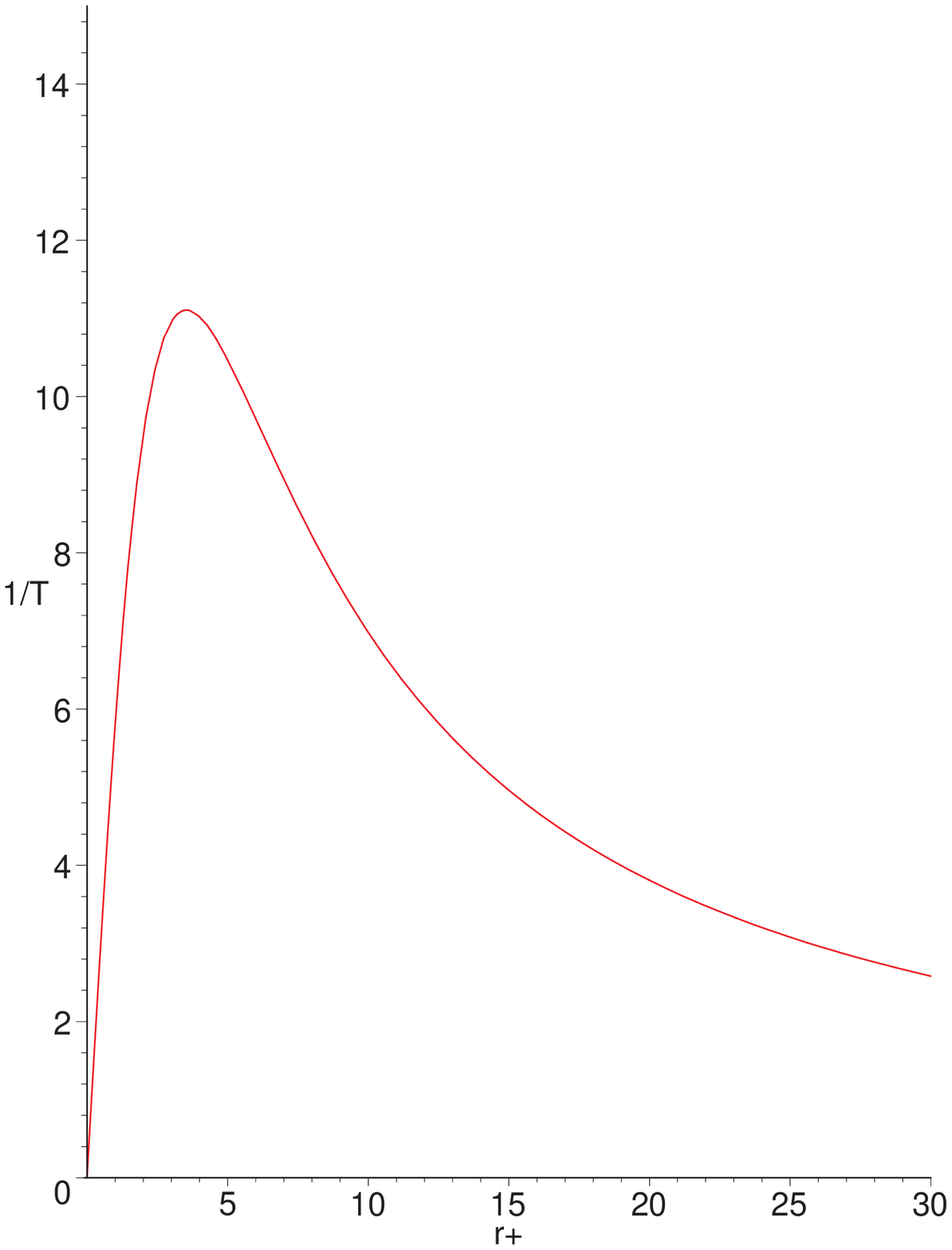,height=3.0in}
\caption{The inverse temperature {\it vs.} horizon radii, for
$q{>}q_{\rm crit}$, $q{<}q_{\rm crit}$, and $q{=}0$,
respectively. $q_{\rm crit}$ is the value of $q$ at which the turning
points of $\beta(r_+)$ appear or disappear.  (The values $n{=}4$,
$l{=}5$ and $q=25,5,0$ have been used here.)  The divergences (here,
shown by the vertical lines at $r_e{=}0.98$ and $4.05$) are at zero
temperature, where the black hole is extremal. The final graph, for
the uncharged case, may be thought of as a limit of the previous
graphs where the divergence disappears, showing that small
Schwarzschild black holes have high temperature.}
\label{betaplot1}
\end{figure}

\noindent
The inverse temperature, $\beta$, is given by eqn.~(\ref{betaform}).
It is useful to plot the temperature as a function of horizon radius
(size of black hole) for future use. There are two basic scales in
this expression for $\tilde I$, set by $q$ and $l$, and so we expect
that there will be two distinct regimes which may display distinct
phase structure: $q{\ge}q_{\rm crit}$ and $q{<}q_{\rm crit}$.  For
comparison, we also show the case of $q{=}0$ (see
fig.~\ref{betaplot1}).

The critical charge $q_{\rm crit}$ is the value of $q$ at which the
turning points of $\beta(r_+)$ appear or disappear. With $q{=}q_{\rm
crit}$, the periodicity $\beta{=}\beta(r_+,q,l)$ will have a point of
inflection with respect to $r_+$ derivatives. Hence we can
simultaneously satisfy
\begin{equation}
{\partial\beta\over\partial r_+}=0={\partial^2\beta\over\partial r_+^2},
\end{equation}
with $r_+{=}r_{\rm crit}$ and $q{=}q_{\rm crit}$. A little algebra then
yields
\begin{equation}
r_{\rm crit}^2={(n-2)^2\over n(n-1)}\,l^2\quad{\rm and}
\qquad
q_{\rm crit}^2 ={1\over(n-1)(2n-3)}\left({(n-2)^2\over n(n-1)}\right)^{n-2}
l^{2n-4}.\label{thecritics}
\end{equation}
Therefore we have for $n{=}3$, $q_{\rm crit}{=}l/6$ and for $n{=}4$,
$q_{\rm crit}{=}l^2/3\sqrt{15}.$

In this case, the figures show that for small charge ({\it i.e.},
below $q_{\rm crit}$), there can be {\sl three} branches of black hole
solutions, to which we will refer later. The middle branch is
unstable\footnote{Its slope is positive and hence its specific heat is
negative: according to eqn.~(\ref{grandc}), ${\partial_\beta S}
\propto r_+^{n-2}{\partial_\beta r_+}.$} while the branch
with the smallest radii is new, and will play an interesting role in
the thermodynamics. For zero charge, we return to the familiar two
branch situation of Schwarzschild, while for large charge, we have a
situation analogous to that seen for the large fixed potential.

\section{Thermodynamics and phase structure}
\label{sec:thermo}

\subsection{Fixed Potential}
This is the grand canonical ensemble, at fixed temperature and fixed
potential. The grand canonical (Gibbs) potential is
$W{=}I/\beta{=}E{-}TS{-}\Phi Q.$ Using the expression in
eqn.~(\ref{actionone}), we may compute the state variables of the
system as follows:
\begin{eqnarray}
E=&& \left(\frac{\partial I}{\partial\beta}\right)_\Phi
-\frac{\Phi}{\beta}\left(\frac{\partial I}{\partial\Phi}\right)_\beta
=\frac{(n-1)\omega_{n-1}}{16\pi G}m=M, \nonumber\\ 
S=&&
\beta\left(\frac{\partial
I}{\partial\beta}\right)_\Phi-I=\frac{\omega_{n-1}r_+^{n-1}}{4G}
=\frac{A_H}{4G},\qquad {\rm and}\nonumber\\
Q=&&-\frac{1}{\beta}\left(\frac{\partial
I}{\partial\Phi}\right)_\beta=
\sqrt{2(n-2)(n-1)}\left(\frac{\omega_{n-1}}{8\pi G}\right)q.
\label{grandc}
\end{eqnarray}
Together, they indeed satisfy the first law: $dE=TdS+\Phi dQ$.

\begin{figure}[ht]
\hskip-.5cm
\psfig{figure=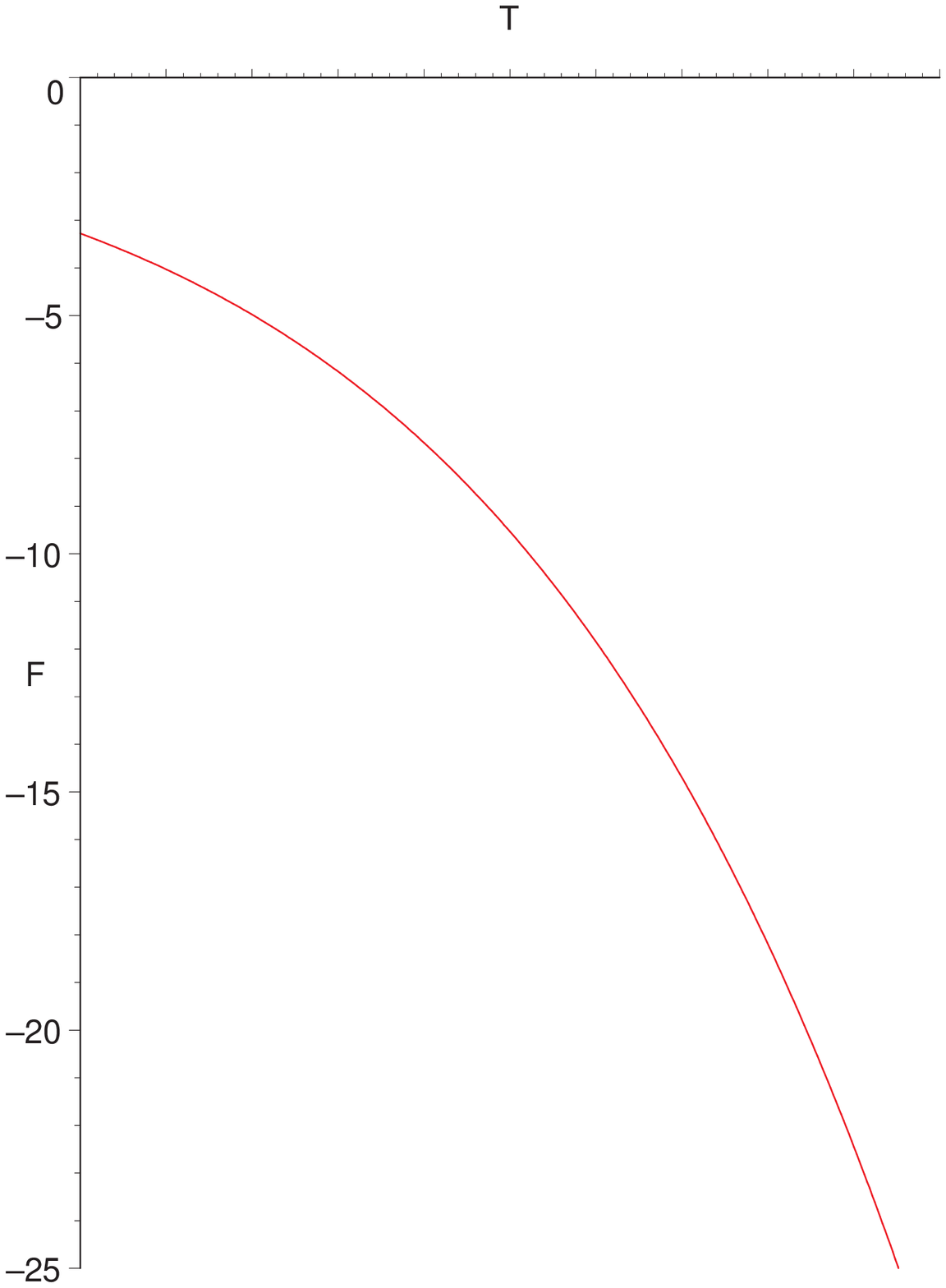,height=3.0in}
\psfig{figure=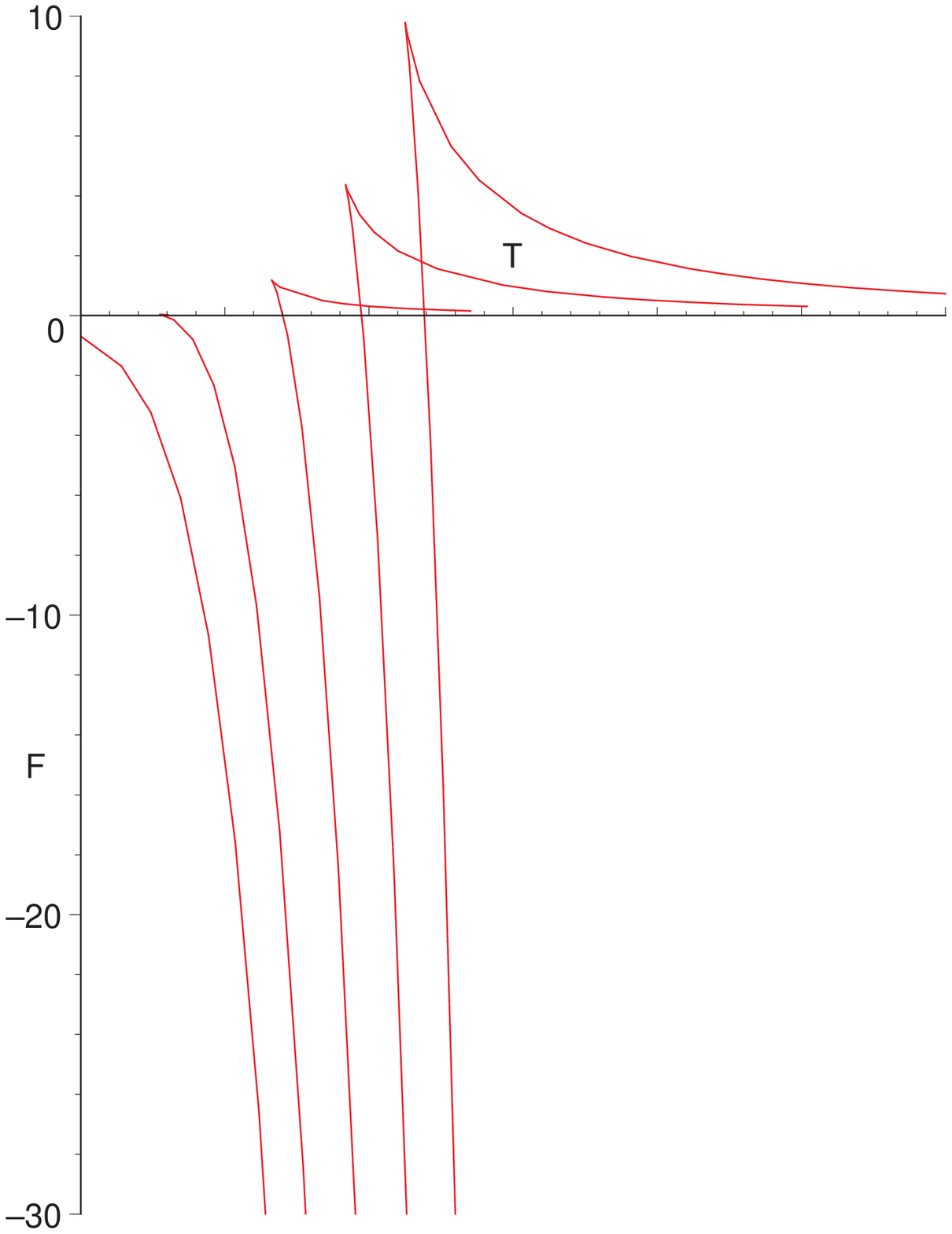,height=3.0in}
\psfig{figure=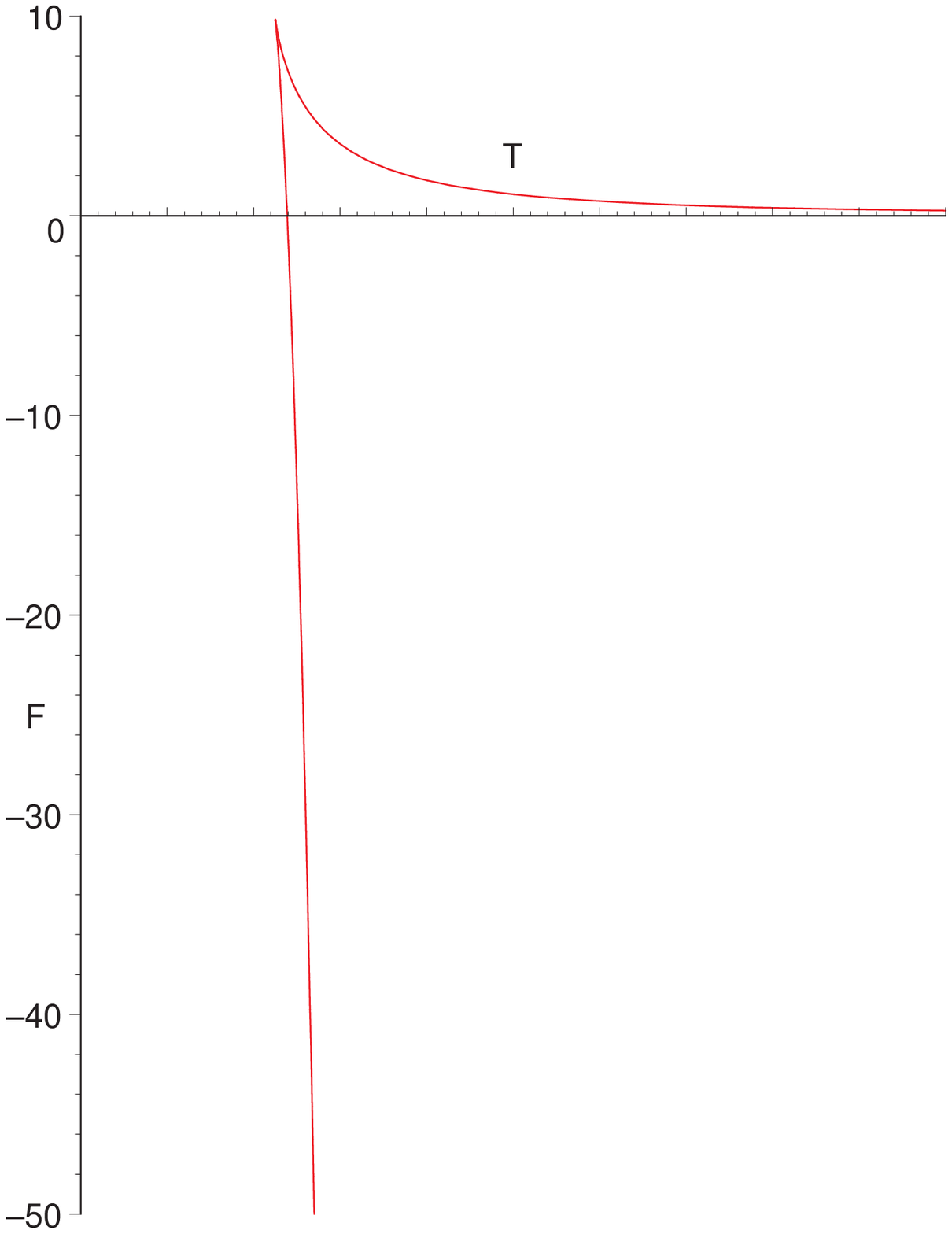,height=3.0in}
\caption{On the left is a graph of the free energy {\it vs.}
temperature for fixed potential ensemble for large $\Phi$. (The values
$n{=}4$, $G{=}1$, $l{=}10$, $\Phi{=}1$ have been used here.) The
center graph depicts a family of free energy curves for different
values of $\Phi$. Note the crossover from the cusp ($\Phi{<}1/c$) to
the single branch ($\Phi{>}1/c$) behaviour. On the right is the free
energy curve for the uncharged (or $\Phi{=}0$) ensemble, showing the
physics familiar from the Schwarzschild case: visible are the two
branches consisting of smaller (unstable) and large (stable) black
holes. The entire unstable branch has positive free energy while the
stable branch's free energy goes (rapidly, on this scale) negative for
all $T{>}T_c$.}
\label{freeplot1}
\end{figure}

In order to study the phase structure and stability, we must observe
the free energy $W{=}I/\beta$ as a function of the temperature. It is
shown in  figure~\ref{freeplot1}.

The interpretation of this is as follows. At any non--zero
temperature, for large potential ($\Phi{>}1/c$) the charged black hole
is thermodynamically preferred, as its free energy (relative to the
background of AdS with a fixed potential) is strictly negative for all
temperatures. 

This behaviour differs sharply from the small potential ($\Phi{<}1/c$)
situation, which is qualitatively the same as the uncharged case: In
that situation, in finite volume, the free energy is positive for some
range $0{<}T{<}T_c$, and it is only above $T_c$ that the
thermodynamics is dominated by Schwarzschild black holes (the larger,
stable branch), after their free energy is negative. (See the center
graphs in figure~\ref{freeplot1}.)

So for high enough temperature in all cases the physics is dominated
by non--extremal black holes. In this case, (after converting
gravitational to field theory quantities\footnote{We do this using the
standard formulae derived from the brane
geometry~\cite{juan,edads,edadsii}: For $n{=}3$, $G{\sim}l^{-7}$ and
$l{\sim}N^{1/6}$; for $n{=}4$, $G{\sim}l^{-5}$ and $l{\sim}N^{1/4}$;
for $n{=}6$, $G{\sim}l^{-4}$ and $l{\sim}N^{1/3}$.})  the free energy
and entropy behave at ultra--high temperature as
\begin{eqnarray}
F&&\sim V_{n-1}T^{n}N^{p(n)}\nonumber \\ S&&\sim V_{n-1}T^{n-1}N^{p(n)}, 
\label{behave}
\end{eqnarray}
where $V_{n-1}$ is the $(n{-}1)$--dimensional spatial volume upon
which the field theory lives.  This is the ``unconfined'' behaviour
appropriate to the dual $n$--dimensional field theory. The function
$p(n)$ is $2$ when $n{=}4$; $3/2$ when $n{=}3$; $2/3$ when
$n{=}6$. The resulting power of $N$ shows how the number of
unconfined degrees of freedom of the theory goes with $N$, by
analogy with the case of $n{=}4$ where $N^2$ counts the dependence on
the number of degrees of freedom on $N$ for an $SU(N)$ gauge theory.

At low temperatures, and for $\Phi{>}1/c$, we have something very
new. Notice that as we go to $T{=}0$, the free energy curve approaches
a maximum value which is less than zero. This implies that even at
zero temperature the thermodynamic ensemble is dominated by a black
hole. From the temperature curve~(\ref{betaplot3}) it is clear that it
is the extremal black hole, with radius $r_+{=}r_e$. For $\Phi{=}1/c$,
at $T{=}0$ we recover AdS space.

So this suggests that even at zero temperature the system prefers to
be in a state with non--zero entropy (given by the area of the black
hole). Notice that this $T{=}0$ situation displays the
``confined'' behaviour characteristic of the ordinary
conformally invariant zero--temperature phase, despite the presence of
the black hole. This follows from the fact that the temporal Wilson
lines will still have zero expectation value, as the fundamental
strings which define them cannot wind the horizon which has infinite
period at zero temperature.  Similarly, spatial Wilson lines will
not display the area law behaviour, because the fundamental string
world--sheets cannot be obstructed by the horizon, because at
extremality, it is infinitely far away down a throat.

Having pointed out this intriguing possible zero temperature
behaviour, we expect that for the case of {\it fixed potential}
considered here, this is not the complete story. We must allow for the
possibility that the extremal black hole might decay due to processes
involving Kaluza--Klein particles charged under the $U(1)$. (See the
discussion near the end of section~I.) This possibility cannot be
discounted because the extremal black hole is not supersymmetric, as
pointed out before, and therefore not guaranteed to be stable by the
supersymmetry algebra. We expect that calculations which include the
effects of charge emission will shift the free energy back to zero,
representing the true, equilibrium situation.  Alternatively if we
consider the action (\ref{actionjackson}) on its own merit outside of
string or M--theory compactifications, it may be regarded as part of a
theory without fundamental charged particles.

The resulting thermodynamic phase structure for the fixed potential
ensemble is summarized in the left diagram of figure~\ref{fig:phases}.

\subsection{Fixed Charge}
We have seen that we may consider a $T{=}0$ background containing an
extremal black hole of charge $Q$. Let us now keep this
charge fixed and allow the potential at infinity to vary.

 This is the canonical ensemble, and the
corresponding thermodynamic potential, the free energy, is
${\tilde I}/\beta{=}F{=}E{-}TS.$
The energy, entropy and electric potential are computed as 
\begin{equation}
E=\left({\partial\tilde I\over\partial\beta}\right)_Q=M-M_e,\quad
S=\beta\left({\partial\tilde
I\over\partial\beta}\right)_Q-{\tilde I}={A_H\over 4G},\quad{\rm and}\quad
\Phi={1\over\beta}\left({\partial\tilde I\over\partial
Q}\right)_\beta= {1\over c}\left({q\over r_+^{n-2}}-{q\over r_e^{n-2}}\right).
\end{equation}
In this case $E$ measures the energy above the ground state, which is
the extremal black hole.  Together, they satisfy the first law, which
in this case should be written as $dE{=}TdS{+}(\Phi{-}\Phi_e)dQ$.

The free energy as a function of temperature is shown below for the
cases of small and large charge, respectively (compare to the third
graph in figure~(\ref{freeplot1}) for the uncharged case).

\begin{figure}[ht]
\hskip-0.5cm
\psfig{figure=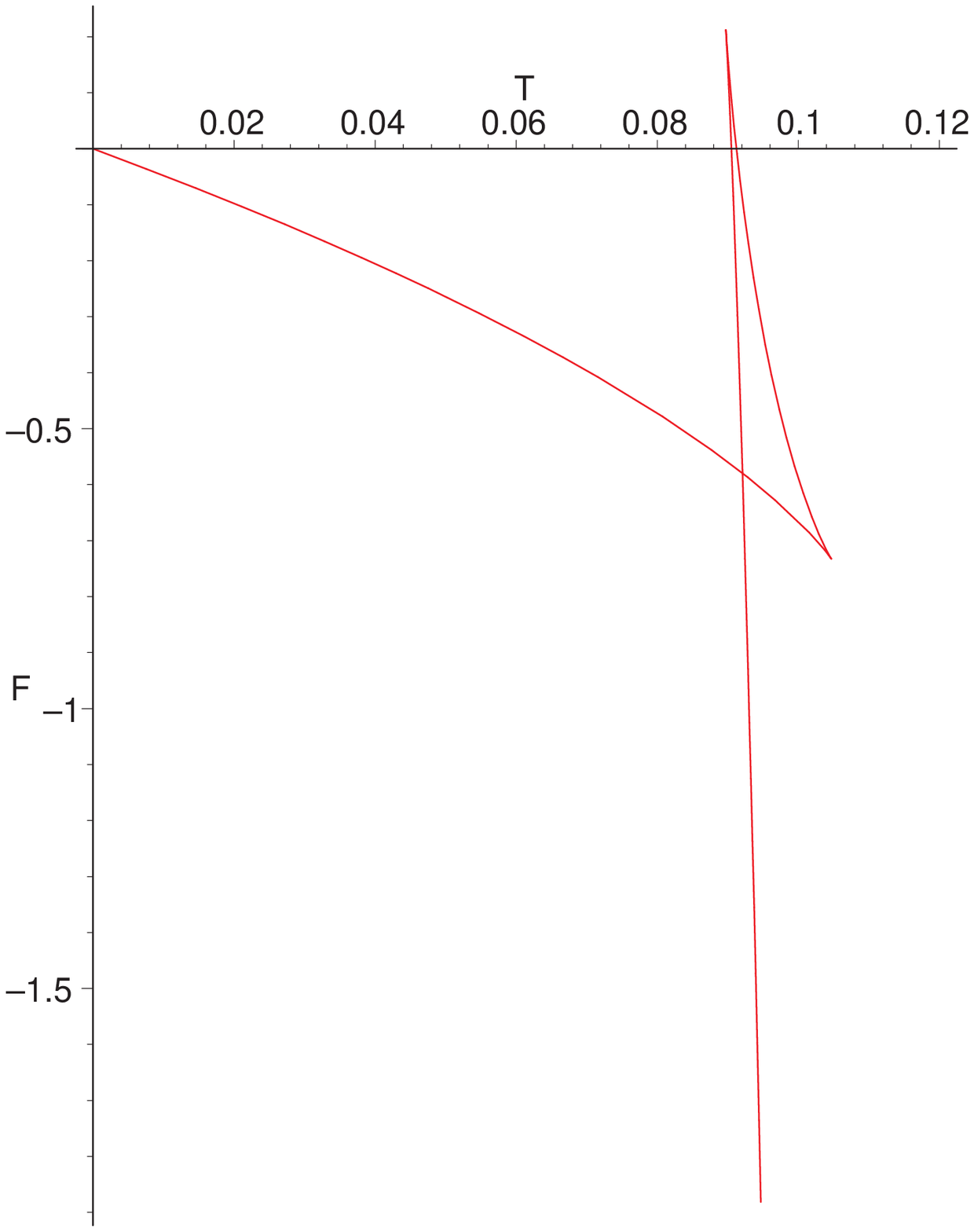,height=3.0in}
\psfig{figure=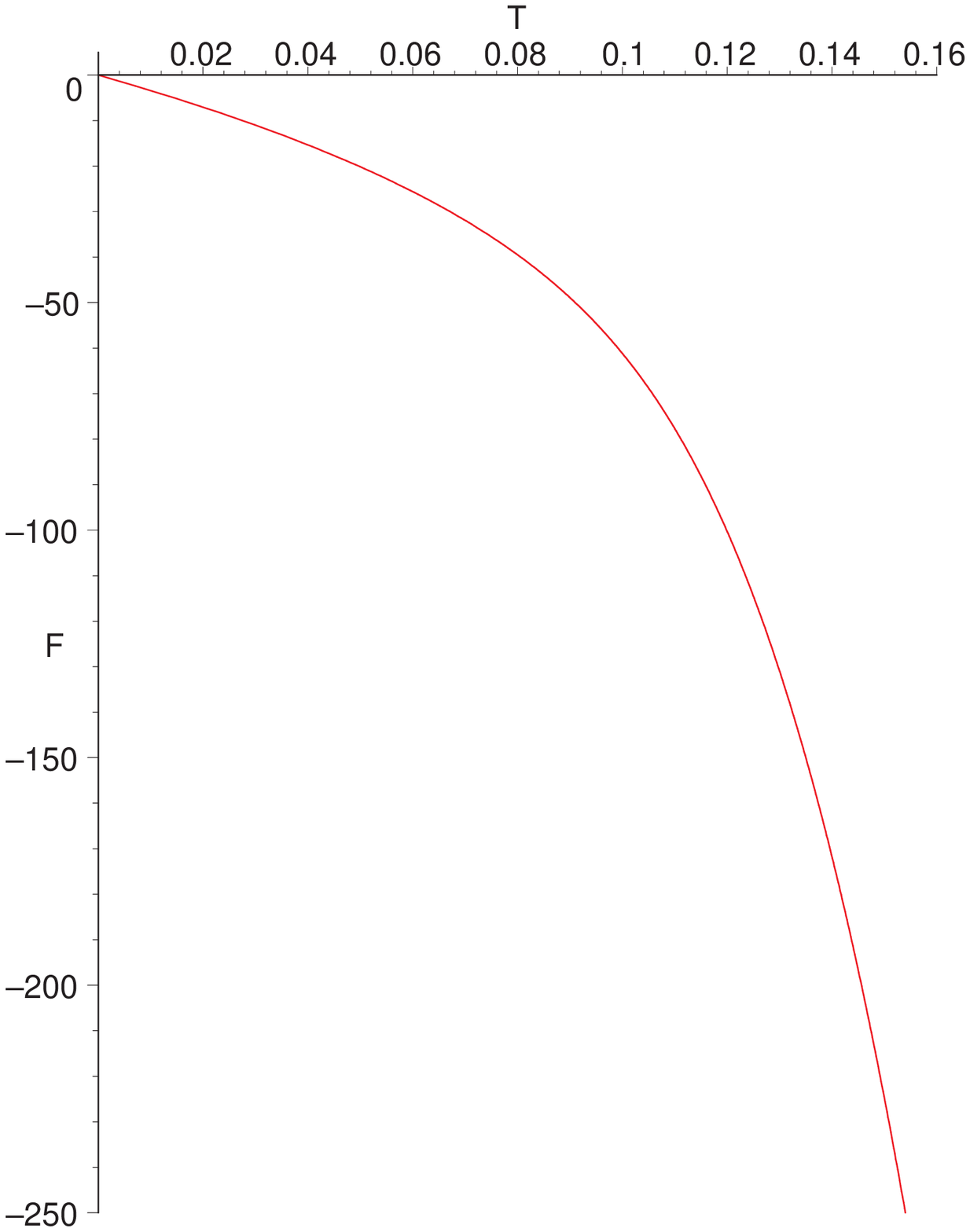,height=3.0in}
\psfig{figure=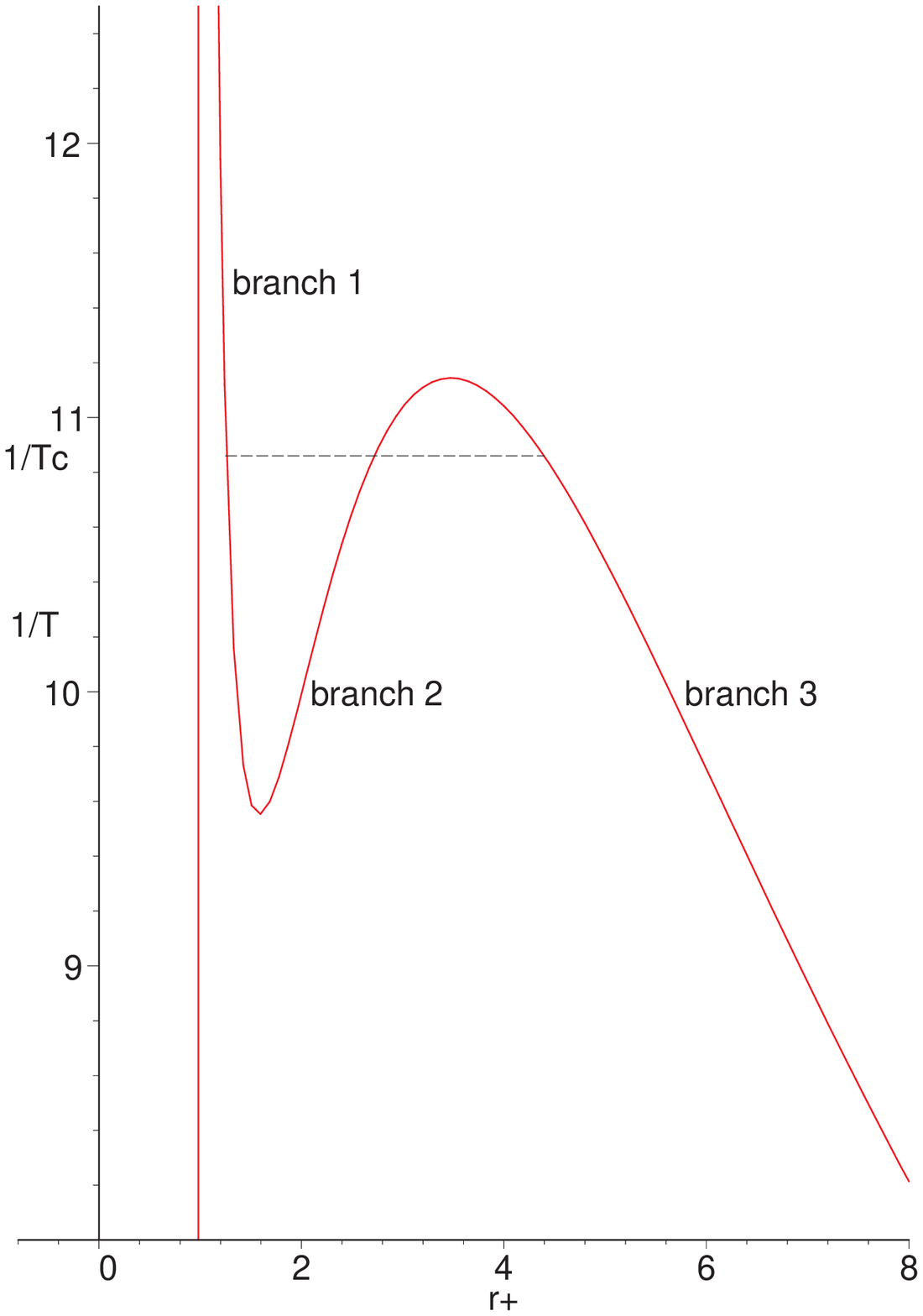,height=3.0in}
\caption{The first two graphs show the free energy {\it vs.}
temperature for the fixed charge ensemble. The situation for
$q{<}q_{\rm crit}$ and $q{\ge}q_{\rm crit}$, respectively, are
plotted. (The values $n{=}4$, $G{=}1$, $l{=}5$ and $q{=}1,25$ have been
used here.)  The first graph is the union of three branches. Branch 1
emanates from the origin, and merges with branch 2 at a cusp. Branch 3
forms a cusp with the other end of branch 2, and continues towards the
bottom right.  The graph on the right shows how the branches arise
from the inverse temperature curves of eqn.~(\ref{betaform}). (See
text for discussion of critical temperature $T_c$.)}
\label{freeplot3}
\end{figure}

That there are three branches for the small charge case follows from
the second graph in figure~\ref{betaplot1}, which is magnified and
labeled in fig.~\ref{freeplot3}, on the right. From there, it is
clear that for low temperature there can only be one solution
(``branch 1'') for the black hole radius. At some temperature
$T_1{=}1/\beta_1$, the origin of two new branches (``branches 2 and 3'')
of solutions appears ($T_1{=}0.089,\beta_1{=}11.15$ for the chosen
parameters in the plot.). Above this temperature (below $\beta_1$),
there are therefore three distinct branches of solution until at
temperature $T_2{=}1/\beta_2$, ($T_2{=}0.105,\beta_2{=}9.55$ in the
plot) two of the branches (1 and 2) coalesce and disappear, leaving
again only a single branch (3), which persists for all higher
temperatures.

Returning to the free energy plot, the meaning is now clear. Starting
to the extreme left of the plot, (low temperature) we see that there
is a single branch of free energy, corresponding to the branch 1
solutions. At $T_1$, branches 2 and 3 appear on the graph and separate
from each other at higher temperatures. At $T_2$, branches 1 and 2
coalesce and disappear, while branch 3 persists for all higher
temperatures, continuing to the left. 

So from zero temperature the negative free energy of branch~1 means
that those non--extreme black holes dominate the thermodynamic
ensemble.  At temperature $T_c$, ($T_c{=}0.092$ in the plot) the free
energy of branch 3 is actually more negative than that of branch~1,
and so that branch of non--extremal black holes takes over the physics
and continue to do so for all higher temperatures.

The situation at $T_c$ is a genuine finite temperature phase
transition, of first order. (Notice from the first graph in
figure~\ref{freeplot3} that the free energy is continuous, but its
first derivative is discontinuous.) This results from the jump (along
the dotted line in the final graph in figure~\ref{freeplot3}) from
branch 1 to branch 2, from small to large $r_+$ black holes, as the
temperature increases. As the entropy is proportional to
$r_+^{n-1}$, there is a jump in the entropy, or a release of ``latent
heat''.

As we approach the critical value, $q_{\rm crit}$, of the charge
representing the crossover into the large charge regime, the kink in
the free energy ---and therefore the transition--- vanishes, as
branches 1 and 3 merge (and branch 2 disappears). The difference in
horizon radii between the two branches,
$\rho_+{=}r^{(3)}_+{-}r^{(1)}_+$, may be thought of as an order
parameter for the transition, as it vanishes above $q_{\rm crit}$,
where the transition goes away.

As noted before in the case of fixed potential ensemble, branches 2
and 3 are the exact analogues of the small and large Schwarzschild
black holes of Hawking and Page\cite{hawkpage}, or the small and large
Taub--bolts discovered in the thermodynamic studies of
ref.~\cite{cejm,hawkhunt}. In those papers,
above a certain temperature $T_1$, there were two allowed solutions at
a given temperature, the smaller (branch 2) being unstable and the
larger (branch 3) being stable, which persists to dominate the
thermodynamics above some critical temperature $T_c$. The existence of
a stable branch 1, and its merger with branch $2$ to disappear at
$T_2$ is a new feature when we add a small fixed charge to the
story. Conversely, if we start from a situation where charge is
present on the black hole but the cosmological constant vanishes, then
we find branches 1 and 2, and it is only when the negative
cosmological constant is turned on that branch 3 appears.

For large charge, there is only a single branch allowed, (see
figure~\ref{freeplot3}, the cusps collide and disappear) and the
associated thermodynamic story is correspondingly simpler. The free
energy shows that the non--extreme charged black holes dominate from
$T{=}0$.

In all cases (large or small $Q$), the ultra high temperature phases
are dominated by a black hole and the free energy and entropy have the
characteristic ``unconfined'' field theory behaviour shown in
eqns.~(\ref{behave}).

One might examine the approach to the critical point more
closely. In particular, consider the behaviour of the specific heat
\begin{equation}
c_q\equiv{\partial M\over\partial T}={\partial M/\partial r_+\over
\partial T/\partial r_+}\ .
\end{equation}
With $q{=}q_{\rm crit}$, as the temperature approaches the critical
value, one finds a singularity with
$c_q{\propto}(T{-}T_c)^{-2/3}$. This behavior may be contrasted with
the $(T{-}T_c)^{-1/2}$ singularity found in ref.\cite{14.3}. The
essential difference is of course that near the critical point we have
a point of inflection with $T{-}T_c \propto(r_+{-}r_{\rm crit})^3$,
while ref.\cite{14.3} considers a minimum with $T{-}T_{c}\propto
(r_+{-}r_{\rm crit})^2$.

The evolution of the free energy of the system as a function of charge
is particularly interesting as one goes from zero charge to large
charge. The single cusp of the uncharged (Schwarzschild) system is
joined by a second cusp which comes in from infinity, forming (with
the original one) a section of the well known ``swallowtail'' shape,
familiar as a bifurcation set or ``catastrophe'' in singularity or
catastrophe theory. The significance of this is discussed in the next
section. As we cross over into the large charge regime at some
critical value of $q$, the cusps merge and the free energy becomes a
simple monotonic function. For completeness, we include a series of
plots showing this evolution. (We do not put them on the same axes, as
we did for the fixed potential case, for the sake of clarity.)

\begin{figure}[ht]
\hskip-0.5cm
\psfig{figure=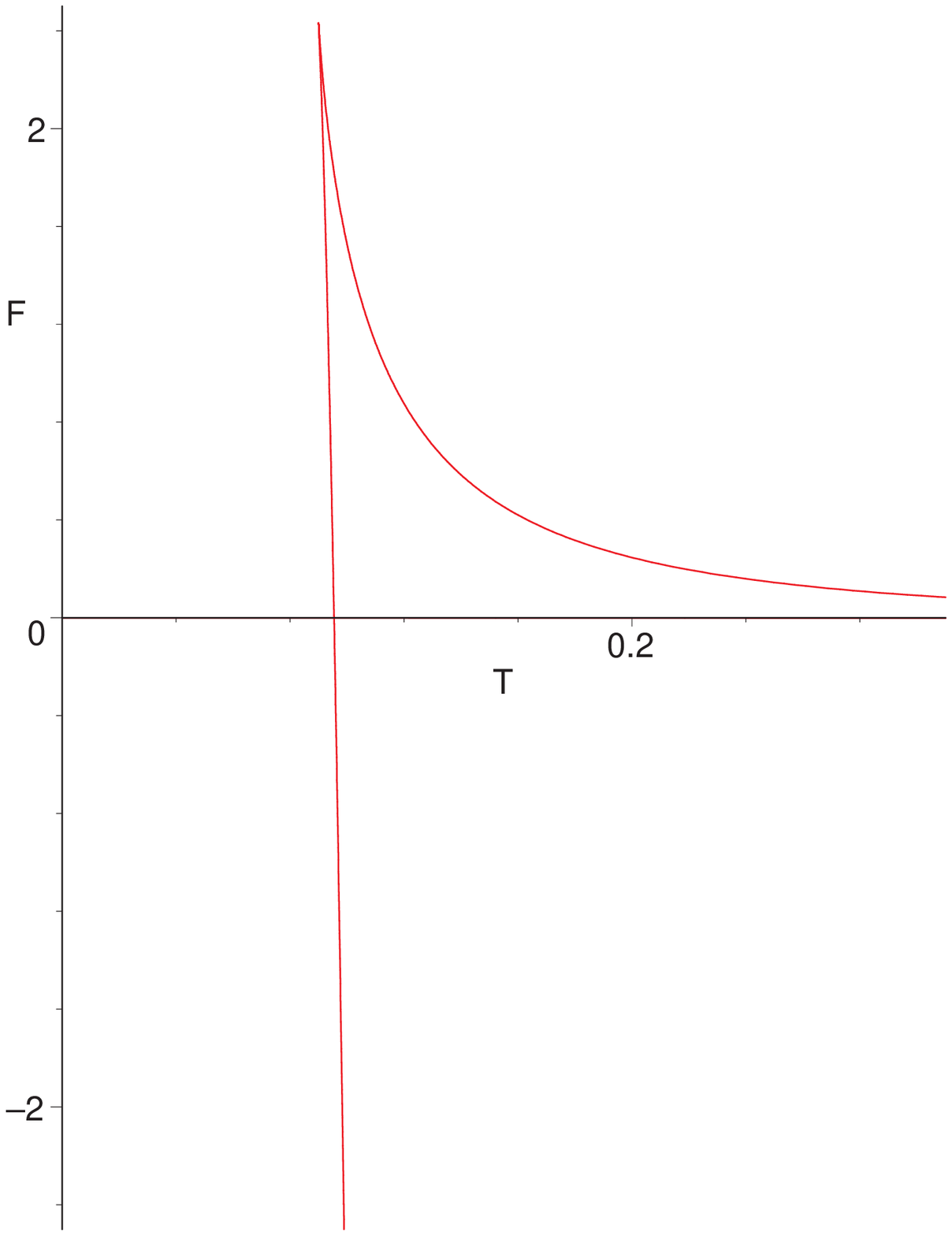,height=3.0in}
\psfig{figure=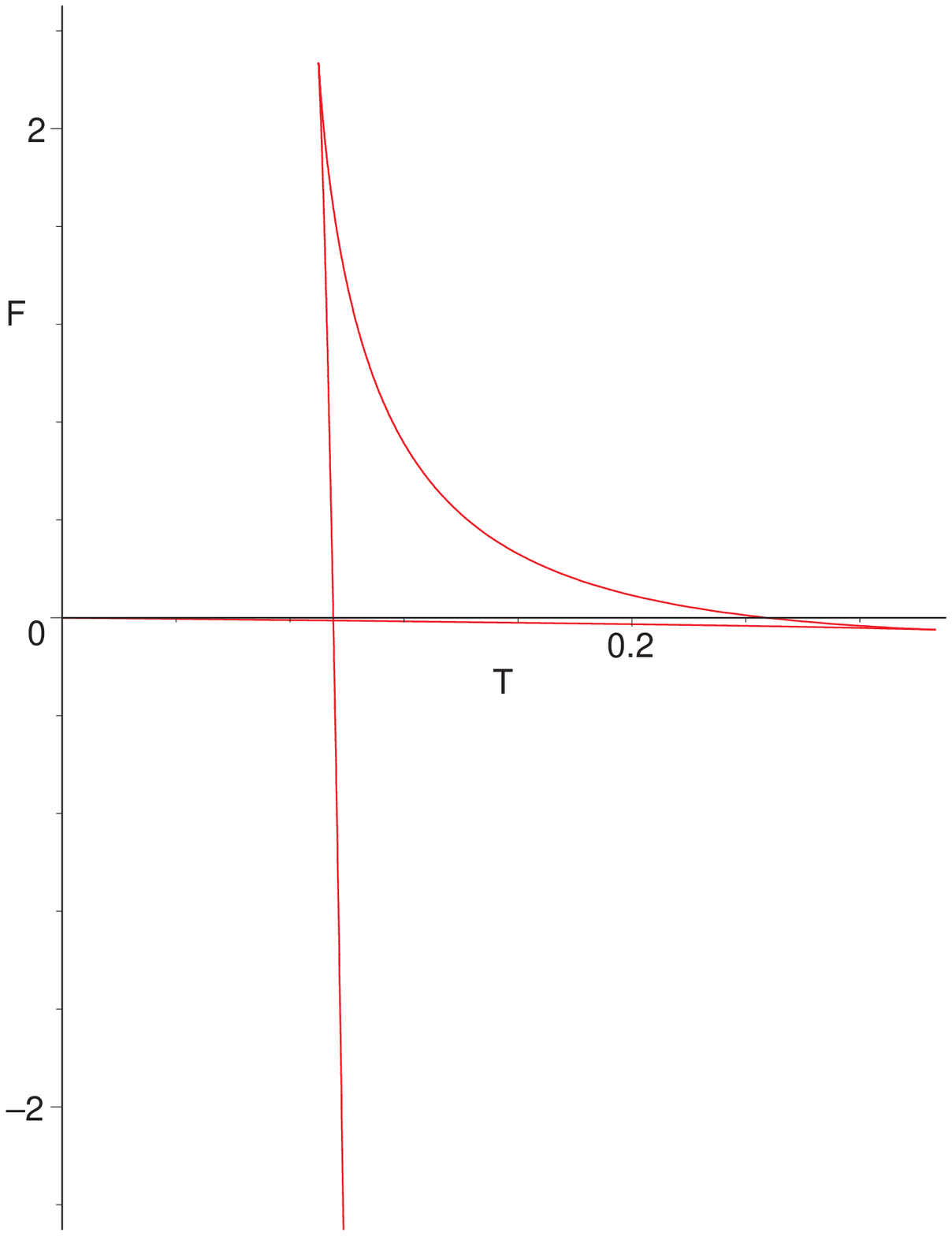,height=3.0in}
\psfig{figure=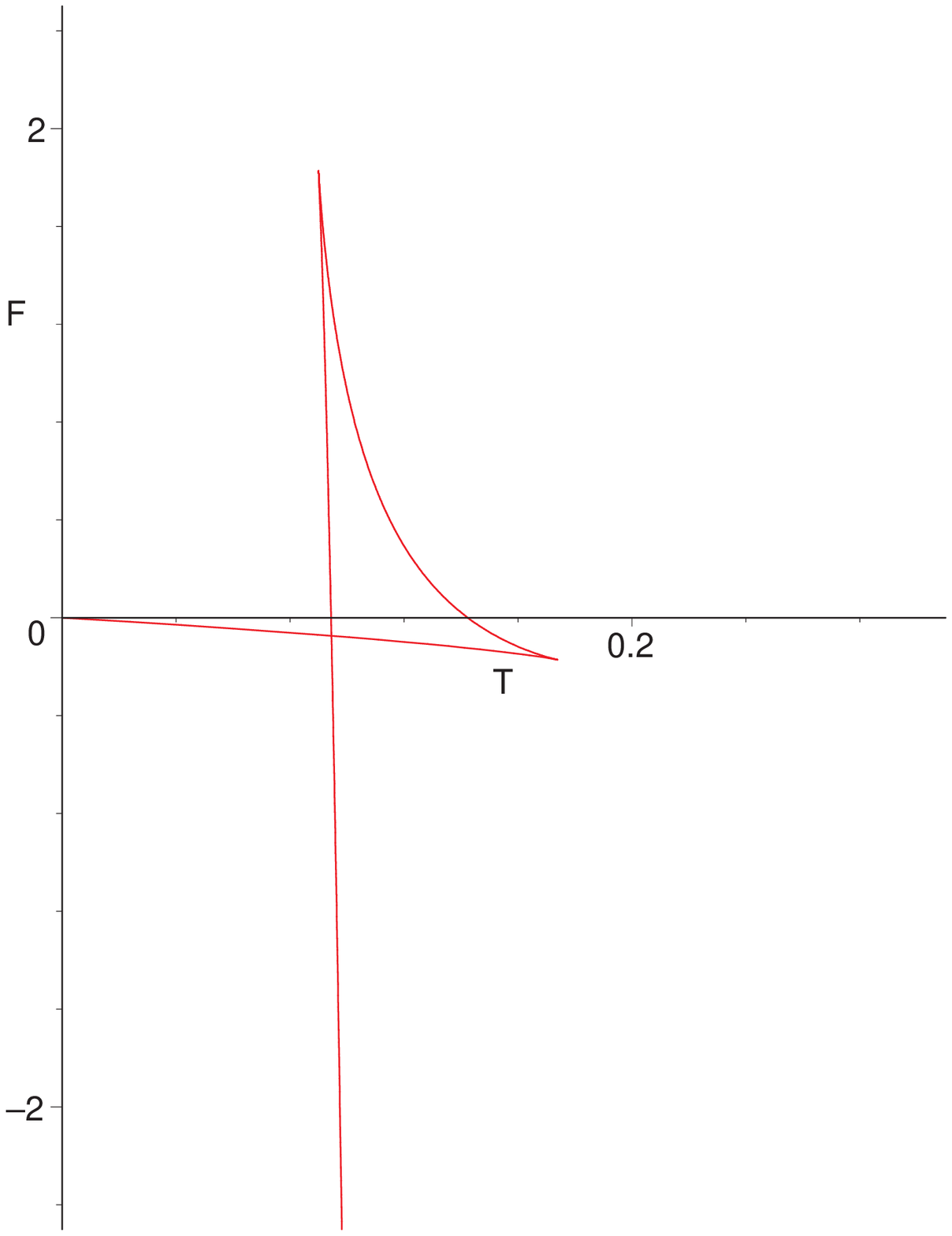,height=3.0in}
\vskip0.5cm
\hskip-0.5cm
\psfig{figure=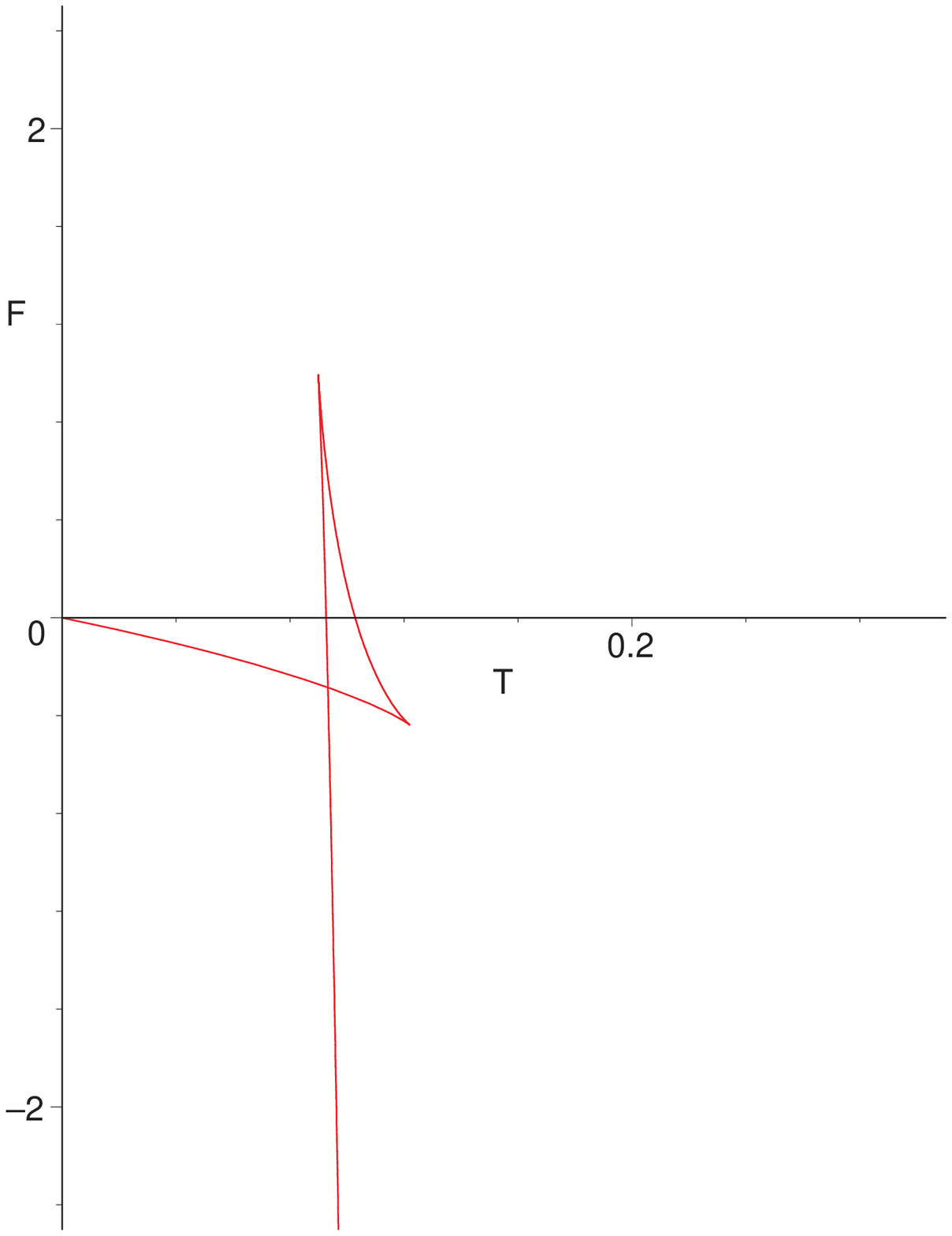,height=3.0in}
\psfig{figure=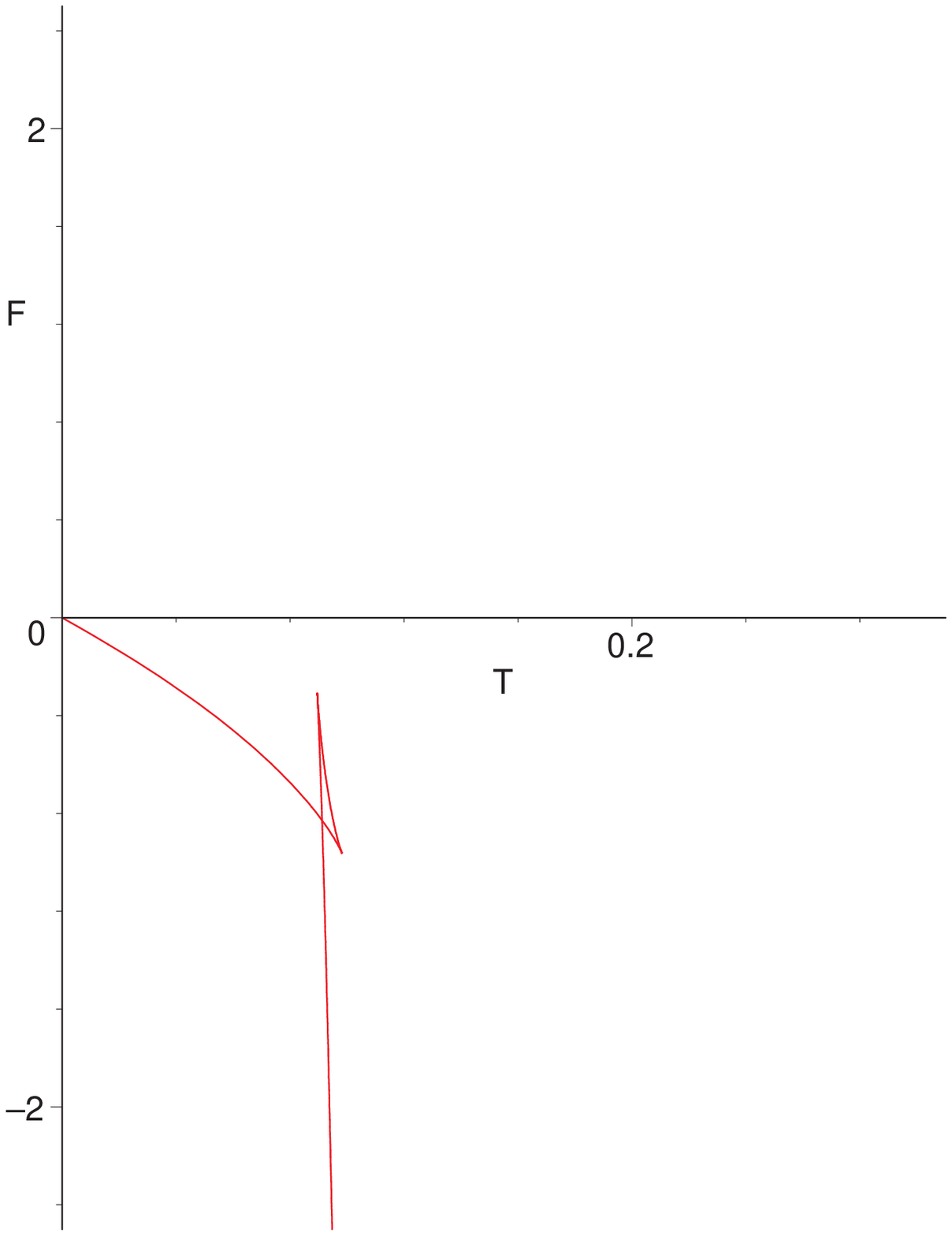,height=3.0in}
\psfig{figure=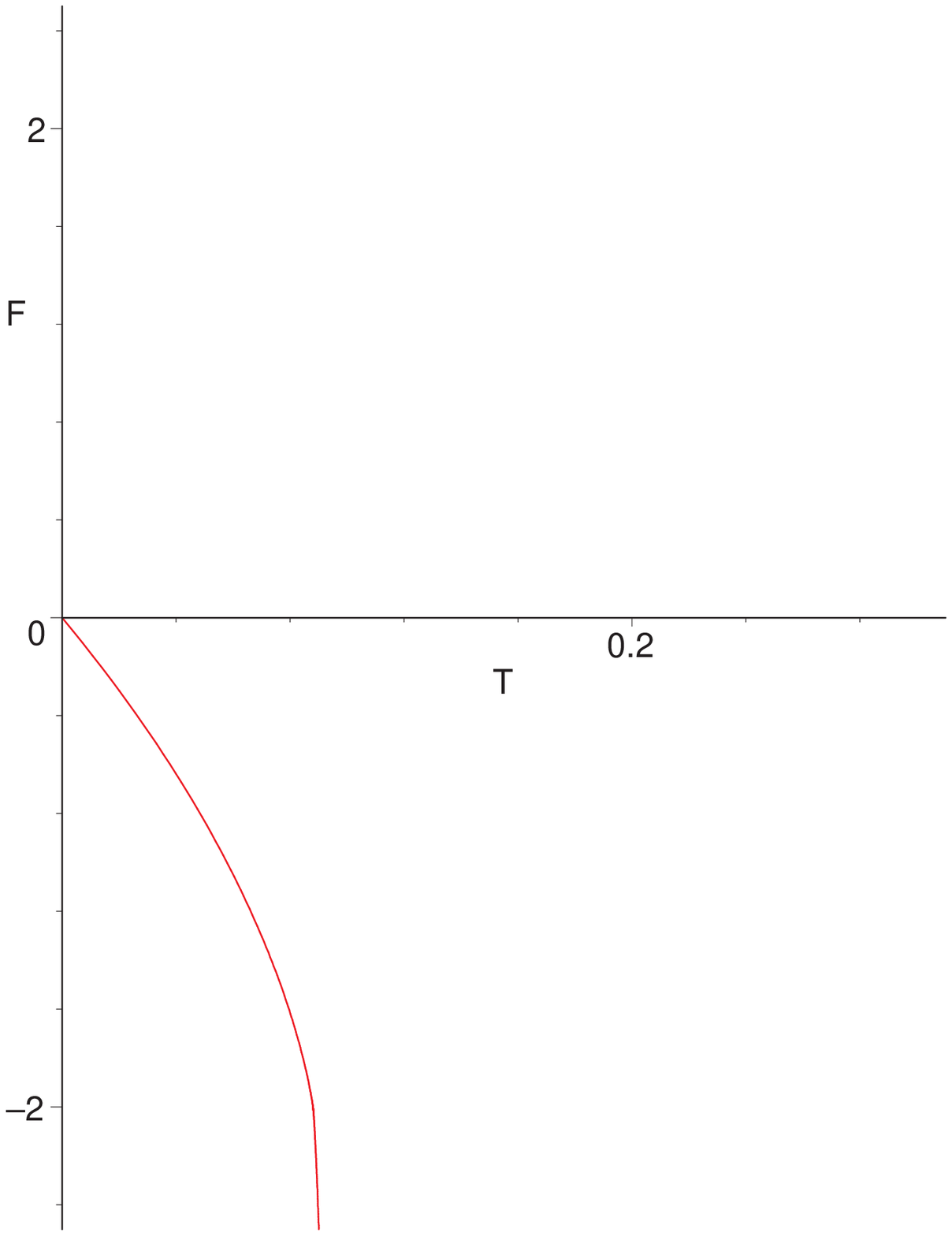,height=3.0in}
\caption{The free energy {\it vs.} temperature for the fixed charge
ensemble, in a series of snapshots for varying charge, starting from
(near) zero charge (top left) and finishing with large charge (bottom
left). The values $l{=}5$, $G{=}1$, and $n{=}4$ are used here. This
complete evolution describes the two dimensional``swallowtail''
catastrophe.}
\label{freeplots}
\end{figure}

The resulting thermodynamic phase structure for the fixed charge
ensemble is summarized in the diagram on the right in
figure~\ref{fig:phases}.

\section{Catastrophic Holography?}
\label{cats}
We cannot refrain  from 
further general comments upon the meaning and structure of the curves
that we have uncovered in the previous sections. Although we plotted
only the cases for the $n{=}4$ case, representing AdS$_5$ (and hence
four dimensional field theory), the same universal structures appear
in the cases $n{=}3$ and 6 as well, giving the same pleasing phase
structure for the fixed charge ensemble.

The phase structure that we uncovered for the fixed charge ensemble
should remind the reader of the classic van der Waals--Maxwell
behaviour, modeling the liquid--gas system. Indeed, they are
isomorphic. The $\beta(r_+)$ curve (the middle graph of
figure~\ref{betaplot1}) should recall the graph of the $P(V)$ van der
Waals equation of state, where $P$ (the pressure) is replaced here by
$\beta$ and $V$ (volume) by $r_+$.

The instability of branch 2 is then simply the familiar instability of
the corresponding section of the van--der Waals curve. The jump from
branch~1 to branch~3 which we deduced from the form of the free energy
is the precise analogue of the Maxwell construction\footnote{Although
one can formulate an adequate ``equal area law'' for this system, here
we have used the lowest free energy condition from which it follows in
the case of the liquid--gas system.}.  In the isomorphism between our
parameters and those of the van der Waals--Maxwell system, our charge
$Q$ is equivalent to their temperature $T$.

The instability of branch~2 in both languages makes intuitive sense:
as one increases the pressure, the volume should {\it decrease}, and
therefore the positively sloped branch is not stable. A similar
statement holds for the black holes after making the translation to
the current situation: For black holes in equilibrium with the heat
bath, an increase in the temperature results in an increase in the
black hole radius and hence mass, for stable black holes. Notice that
this also follows from the first law, recalling that the entropy is a
positive power of the radius. So the positive slope branch of the
$\beta(r_+)$ curve is generally unstable.

In the language of catastrophe theory\cite{catastrophe} ---the study
of jumps in some ``state variables'' as a result of smooth changes in
``control variables''--- the physical solutions of the $\beta(r_+,q)$
curve, viewed as a two dimensional surface in $(\beta,q,r_+)$ space,
is the ``control surface'' of the ``cusp catastrophe''. The cusp shape
is the union of points in the $(\beta,r_+)$ plane (the control
variables) where the state variable (the allowed value of $r_+$) jumps
from branch~1 to branch~3, as branch~2 is unstable.  After applying
the minimum free energy condition to determine the allowed branches
(the ``Maxwell criterion''), the cusp catastrophe appears in the
$(q,\beta)$ plane, (or equivalently the $(Q,T)$ plane) collapsed to
the critical line (see figure~\ref{fig:phases}) (or ``vapour pressure
curve'') along which the two types of black hole can coexist, and
across which there is a phase transition. The end of the line, at the
critical value $q_{\rm crit}$, where branch~2 disappears, is the point
where the distinction between branch~1 and~3 goes away.  The order
parameter, $\rho_+$ for this critical point is the radius difference
of the branches $\rho_+{\equiv}r^{(3)}_+{-}r^{(1)}_+$.  Beyond the
critical charge there is no phase transition ($\rho_+{=}0$) in going
from branch~1 black holes to branch~3 by increasing the
temperature. This is of course the familiar statement that above a
critical temperature, there is no phase transition in going from a gas
to a liquid by increase of pressure.

Intriguing is the fact that the two dimensional free energy surface
$F(\beta,Q)$ forms the shape of the swallowtail catastrophe (see
figure~\ref{freeplots}). (Note that for $n{=}3,4$ and $6$ the shape is
the same.) This naturally follows from the ability of the $\beta(r_+)$
curve to produce three branches, and the resulting shape for the free
energy curve is the union of three branches. 

Here, the swallowtail does not have the usual interpretation as a
bifurcation surface (like the cusp does above) but it is natural to
wonder whether its appearance tells us that there is some universality
at work here.  This is because the language of catastrophe theory is
largely a classification of the possible distinct types of bifurcation
shapes that can occur. This classification (which, for the common
``elementary'' cases is of A--D--E type) is equivalent to the (perhaps
more familiar) classfication of singularities\cite{arnold}. A natural
question is whether or not the inclusion of more control parameters
will always result in a free energy curve of a shape (and
corresponding phase structure) which falls into the classification. It
would certainly be amusing to find yet another case of the A--D--E
classification appearing in string and M--theory physics.

\section{Concluding Remarks}
\label{conclude}
The study of the thermodynamics of black holes in
Einstein--Maxwell--anti--deSitter is highly relevant to the
thermodynamics of certain superconformal field theories with a
background global current switched on. This follows from the logic of
the AdS/CFT correspondence, and the fact that the EMadS system can
arise as the near--horizon physics of rotating M2-- and D3--branes,
and it should therefore be regarded as the effective theory of the
strongly coupled field theory living on the rotating branes'
world--volume\footnote{Strictly speaking, in performing a
near--horizon limit explicitly on a brane solution, one gets the
infinite volume limit black hole solutions of EMadS, but the
interpretation of the finite volume solutions clearly follows.}.

The phase structures of the charged black hole systems studied here,
and summarized in figure~\ref{fig:phases}, are markedly different from
those of the uncharged systems studied before in this
context~\cite{edads,edadsii,hawkpage,cejm}. The addition of charge
revealed a rich phase structure, with precise analogues to classic
thermodynamic systems. The physics is consistent with a dual field
theory interpretation.

In all cases, the infinite volume limit can be found by taking the
limits given in eqn.~(\ref{scaled}). This scaling may be applied to the
expressions for the actions (eqns.~(\ref{actionone})
and~(\ref{actiontwo})) and the period (eqns.~(\ref{betaform}) and
(\ref{betaformtwo})). In all cases, the result is that there is only
one branch of black hole solutions (like the large charge and
potential situations had in finite volume), and the free energy is
negative definite, showing that the thermodynamics is dominated by
black holes for all temperatures. Of course, this is what we should
expect, from the field theory point of view.

As we commented before, the gauge field in the AdS space naturally
couples to a CFT current $J_\mu$, following the prescription
of ref.~\cite{edads}. From the asymptotic variation of the gauge
field~(\ref{pure}) or its corresponding field strength, one then has an
expectation value $\langle J_t\rangle{\sim}q$. Thus one might think of
the CFT state as containing a plasma of (globally) charged quanta.
The precise nature of the CFT state depends on the ensemble, which we
were studying. For the case of the fixed potential, the dual statement
is that a chemical potential conjugate to the global charge has been
introduced leading to the expectation value.  The fixed charge
calculations correspond to an ensemble of CFT states with a fixed
global charge. Thus the difference between the two calculations is
analogous to that between the canonical (fixed $T$) and microcanonical
(fixed $E$) ensembles.

In the context of D3 branes with $n{=}4$, the $SO(6)$ gauge fields
 couple to the R--symmetry currents in the super--Yang--Mills theory.
This aspect of the duality has been used to great advantage to
produce nontrivial consistency tests by comparing correlators
protected by supersymmetry\cite{mm}. Of course in the present case,
with the truncation to EMadS theory, we are focussed on a particular
diagonal $U(1)$ generator of the $SO(6)$ symmetry.

In this context, we can translate the results of the supergravity
calculations to quantitative statements about the strong coupling
behavior of the super--Yang--Mills theory. Up to numerical factors, we
have as usual\cite{juan}: $g^2_{YM}{\sim} g$ and $(l/l_s)^4{\sim}gN$,
(where $g$ is the type~IIB string coupling) as well as
$G_5{\sim}g^2l_s^8/l^5$. It remains fix how the black hole charge
should be characterized in the CFT. The most natural approach is to
measure the physical charge~(\ref{thecharge}) in terms of the
fundamental charge of the Kaluza--Klein excitations in the AdS space,
{\it i.e.}, with $Q{=}{\bar Q}/l$. In the field theory then,
${\bar\rho}{=}\bar Q/V_3$ (where $V_3$ is the spatial volume of the
field theory) essentially counts the number of fundamentally charged
quanta per unit volume in a given state.  Given this framework, we can
consider the field theory content of our results. For example, one
might wonder what the critical charge (\ref{thecritics}) appearing in
the fixed charge phase diagram corresponds to:
\begin{equation}
{\bar Q}_{\rm crit}\sim {l\,q_{\rm crit}\over G_5}\sim N^2.
\end{equation}

In general, translating the entropy, mass or free energy to a field
theory expression produces a complicated function of both the
temperature $T$ and the charge $\bar Q$. One relatively simple case is
the high temperature limit, where the charge essentially plays no role
(see eqn.~(\ref{behave})). Another interesting case to consider is
that of the extremal black holes for which $T{=}0$. By demanding that
$V(r_+){=}0$ and $(\partial V/\partial r)(r_+){=}0$ have a consistent
solution, one finds that the mass and charge parameters are related by
the following expression:
\begin{equation}
\sqrt{z^2-y^2}=(1+z)-\sqrt{1+z},
\end{equation}
where $z{=}3m/l^2$ and $y^2{=}27q^2/l^4$. A simple case to consider
is that of a large black hole with $m{>>}l^2$, for which
$z^3{\simeq}y^4/4$. Further in this limit, one has that $m{\sim}r_+^2$
and so
\begin{equation}
S\sim {r_+^3\over G_5}\sim {l\,q\over G_5}\sim V_3{\bar \rho}\sim
{\bar Q}.
\end{equation}
Notice that implicitly here we are considering a regime where ${\bar
Q}{>>}N$. The lack of dependence of the entropy on $N$ is a signal of
confined behaviour at zero temperature, despite the presence of the
black hole. It would certainly be interesting if this entropy result
could be recovered by considering partitioning of the charge $\bar Q$
amongst the charged excitations of the CFT.

We have left aside the case of compactification of six dimensional
supergravity on $S^3$ to get AdS$_3$. By setting the $S^3$ in rotation
in its two independent rotation planes, in a symmetric fashion, we get
an electric potential in AdS$_3$. Doing so, notice that if we start
from the solution describing a rotating six dimensional black string
(such as the one obtained from the D1--D5 bound state), then, in the
throat limit the rotation of the $S^3$ can be undone by a
diffeomorphism\cite{cc}. In other words, the effective gauge field in three
dimensions is pure gauge. Nevertheless, as shown in
ref.\cite{btz}, there do exist charged black hole solutions in EMadS
theory in three dimensions. These have an electric potential that
diverges logarithmically at infinity, which prevents one from defining
the ensemble at fixed potential.  Nevertheless, if the extremal black
hole background is subtracted, then the fixed charge ensemble can be
appropriately defined. For non--rotating black holes (in
ref.\cite{btz}, the full Kerr--Newman solution is constructed) there is
only one branch, just like we have found for large fixed charge (see
figure~\ref{betaplot1}, left), with corresponding simple thermodynamic
structure given by figure~\ref{freeplot1} (left).

Finally, it is also worth remarking that the close similarity that we
have observed with familiar structures from equilibrium
thermodynamics, and expectations from a dual field theory is further
encouragement (for those who need it) that the quantum mechanics of
black holes is not unlike that of other situations.


\section*{Acknowledgments}
AC is supported by Pembroke College, Cambridge.  RE is supported by
EPSRC through grant GR/L38158 (UK), and by grant UPV 063.310--EB187/98
(Spain).  Support for CVJ's research, and other support of this
project, was provided by an NSF Career grant, \# PHY9733173 (UK).
RCM's research was supported by NSERC (Canada) and Fonds FCAR du
Qu\'ebec.  This paper is report \#'s DAMTP--1999--29, EHU--FT/9902,
DTP--99/9, UK/99--02 and McGill/99--07. We would like to thank Neil
Constable, Mike Crescimanno, Susan Gardner, Harry Lam, Guy Moore,
Malcolm Perry, and Alfred Shapere for comments and useful
conversations. AC, RE and RCM would like to thank the Department of
Physics and Astronomy of the University of Kentucky for its
hospitality, and we would all like to thank the staff of the
William~T.~Young library (University of Kentucky) for the use of
excellent meeting facilities and other services.

\end{document}